\documentclass[aps, prb, twocolumn, superscriptaddress, floatfix]{revtex4-2} 
\usepackage{graphicx, float}
\usepackage[utf8]{inputenc}
\usepackage{mathtools, placeins}
\usepackage{amssymb, amsmath, commath, braket, physics, color}
\usepackage{hyperref}

\hypersetup{
linkcolor=red,
filecolor=blue,
urlcolor=magenta,
}
\usepackage{placeins, standalone}

\newcommand{\Eth}{\ensuremath{E_{\textrm{Th}}}}	
\newcommand{\Ecal}{\ensuremath{\mathcal{E}}}	
\newcommand{\Egap}{\ensuremath{\Delta\Ecal}}	

\newcommand\nPr[2][^n]{\prescript{#1\mkern-2.5mu}{}P_{#2}}

\newcommand\gEM{ \ensuremath{\gamma_\mathrm{E}} }

\newcommand\kTh{ \ensuremath{k_\mathrm{Th}} }

\newcommand\lEV{ \ensuremath{\mu_{N}} }

\newcommand\mean[1]{\ensuremath{\left\langle #1 \right\rangle}}

\newcommand{\ot}{\ensuremath{\mathbb{O}}}

\newcommand{\prob}[1]{\ensuremath{\textrm{P}\del{#1}}}
\newcommand{\pGOE}[1]{\ensuremath{\textrm{P}_{\textrm{GOE}}\del{#1}}}
\newcommand{\pPsn}[1]{\ensuremath{\textrm{P}_{\textrm{Poisson}}\del{#1}}}

\newcommand\sEV{ \ensuremath{\eta_{N}} }
\newcommand\sgE{ \ensuremath{\sigma_E} }

\newcommand\iiserk{Department of Physical Sciences, Indian Institute of Science Education and Research Kolkata, Mohanpur 741246, India}
\newcommand\lboro{School of Science, Loughborough University, Loughborough, Leicestershire LE11 3TU, United Kingdom}

\begin{document}

\title{Signatures of Nonergodicity in Sparse Random Matrices}
\author{Sagnik Seth}\email{ss22ms026@iiserkol.ac.in}
\affiliation{\iiserk}
\author{Adway Kumar Das}\email{A.K.Das@lboro.ac.uk}
\affiliation{\lboro}
\author{Anandamohan Ghosh}\email{anandamohan@iiserkol.ac.in}
\affiliation{\iiserk}

\begin{abstract}
The prevalence of sparsity in the Fock space graph of interacting many-body systems motivates an investigation into the spectral statistics of sparse random matrices with on-site disorder. We numerically determine the delocalization-localization transition in the ground state as a function of the sparsity. The short-range energy correlation in the bulk indicates that the Anderson transition at infinite temperature occurs at the critical percolation limit of the sparse graph. By analytically deriving the energy moments and calculating the shifted kurtosis, we show that the critical sparsity threshold matches the Anderson transition. Furthermore, long-range energy correlations in the bulk spectrum reveal a Thouless energy scale, suggesting a broad nonergodic regime within the delocalized phase.
\end{abstract}

\maketitle
\section{Introduction}
\begin{figure}[t]
    \centering
    \includegraphics[width=\columnwidth]{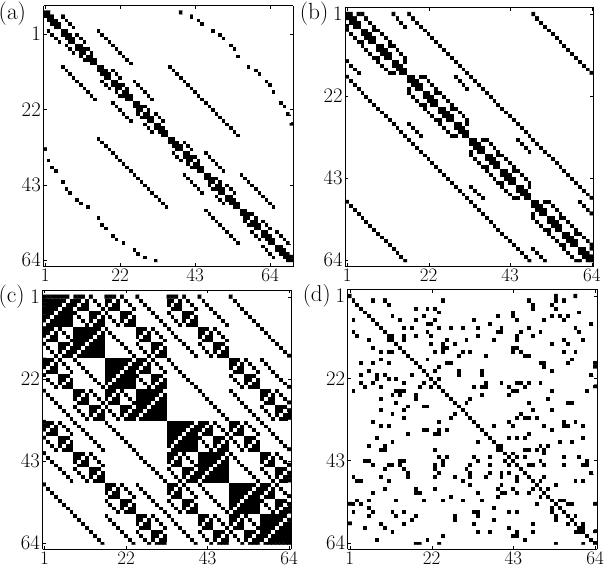}
    \caption[Matrix representations of various sparse Hamiltonian]{{\bfseries Matrix representations of various sparse Hamiltonians:} (a)~1D Heisenberg model with nearest neighbor interactions and periodic boundary condition in the zero magnetization sector for 8 spin-$\frac{1}{2}$ particles in the $\hat{\sigma}^z$ basis~\cite{Das2025}.
    (b)~Anderson model on $4\times 4\times 4$ cubic lattice with nearest neighbor hopping and periodic boundary condition in the site basis~\cite{Lydzba2021, Altshuler1988}.
    (c)~Transverse field Ising model with long-range interaction in the odd-parity sector for 7 spin-$\frac{1}{2}$ particles in the $\hat{\sigma}^z$ basis~\cite{Das2025}.
    (d)~Adjacency matrix of an Erd\H{o}s-R\'enyi-Gilbert graph on 64 nodes with {self-loops where the probability of finding a non-zero edge is 0.1~\cite{Gilbert1959}.}
    }
    \label{fig:graph}
\end{figure}

Recent experimental advances in cold atoms~\cite{Kaufman2016}, ion traps~\cite{Granet2025, Richerme2014}, superconducting qubits~\cite{Dong2025}, nuclear magnetic resonance~\cite{Wei2018} have provided a fresh impetus to explore interacting many-body systems. In the presence of impurities, ergodicity often breaks down and a many-body localized (MBL) phase is predicted in the bulk energy spectrum~\cite{Basko2006, Pal2010, Nandkishore2015, Serbyn2016, Chertkov2021, VallejoFabila2024, VallejoFabila2025, Das2026Arxiv, Suntajs2022}. Such an ergodic-MBL phase transition is reflected in the statistical properties of the matrix representation of a many-body Hamiltonian which is highly sparse and analogous to a single-particle tight-binding model on a graph~\cite{Altshuler1997, Tarzia2020, Roy2020a}. 
In Fig.~\ref{fig:graph}, we show the Fock space graph for prototypical many-body systems like the Heisenberg and transverse field Ising models. Such hierarchical tree-like structures motivated studies on random regular graphs with on-site disorder which exhibit various interesting phenomena, e.g.~Anderson delocalization-localization transition, breaking of ergodicity, sub-diffusive transport and violation of thermalization~\cite{Kravtsov2018, DeTomasi2020, Saha2026Arxiv}.

A typical signature of a MBL eigenstate is the multifractality over the Fock space~\cite{Mace2019, DeTomasi2021}. Thus, it is imperative to study non-regular graphs as opposed to random regular graphs where multifractality is absent~\cite{Monthus2011, Sonner2017, Nosov2022}. 
A prominent example of non-regular graph is the Erd\H{o}s-R\'enyi-Gilbert (ERG) graph, whose adjacency matrix is shown in Fig.~\ref{fig:graph}(d) for the number of nodes $N = 64$ and  the probability of finding a nonzero edge~$p = 0.1$~\cite{Erdos1959, Gilbert1959}. While there is no hierarchical structure in the ERG graph, permutations of the graph nodes~\cite{Cuthill1969} reveal structures similar to those observed in the Fock space representation of many-body systems. Nevertheless, the common feature of all the graphs in Fig.~\ref{fig:graph} is sparsity, which alone can drive delocalization-localization transition.

For the ERG graph, there is a critical percolation limit $p = N^{-1}$ below which the
graph is fragmented and above which there exists a giant connected component with a heterogeneous locally tree-like structure leading to a mobility edge separating localized and extended states in the energy spectrum~\cite{Alt2021}.
Furthermore, random weights can be assigned to graph edges to generate a sparse or dilute random matrix model~\cite{Cugliandolo2024, Khorunzhy1998a} important for understanding various systems like dilute magnets~\cite{Rodgers1988}, mean-field theory of spin glass~\cite{Fyodorov1991, Roy2020, Cicuta2018}.

The spectral properties of the sparse random matrices have drawn considerable interests over the last three decades. The density of states (DOS) of a quantum system can be obtained from the Laplace transform of the complex partition function, which has been studied using the replica method in case of the sparse matrices~\cite{Rodgers1988, Dean2002, Fortin2005, Kuhn2008, Rogers2008, Susca2021}. For $p \to 1$, the DOS follows the Wigner's semicircle law while a tail 
is observed for $p < 1$ and finite $N$~\cite{Khorunzhy1997, Khorunzhy1998, Khorunzhy1998a}. For $p > N^{-1}$, the spectral properties of the giant connected component with random hopping terms and no self-loop is systematically studied in Refs.~\cite{Cugliandolo2024, Baron2026Arxiv}. The absence of self-loop leads to extensively many degenerate eigenstates at zero energy, which are localized on the leaves of the connected component, leading to a divergence in the bulk DOS~\cite{Tapias2023}.

A power-law tail in the local density of states~(LDOS) indicates a weak multifractality of the bulk states at non-zero energy where the fractal dimension $D_1$ is unity, i.e.~the support set of the eigenstate intensities spans a finite fraction of all the nodes but $D_q < 1$ is $q$-dependent for some $q > q_\star$, i.e.~a vanishingly small number of nodes support anomalously large eigenstate intensities~\cite{Silva2022, Silva2025, Baron2026Arxiv}. In contrast, Anderson localized states exist at low energies which are separated from the multifractal extended states by a mobility edge~\cite{Mirlin1991, Biroli1999, Semerjian2002}.

The heterogeneous spectral properties of the weighted adjacency matrix are manifested in the non-equilibrium dynamics on ERG graphs. Particularly, the survival probability of an initially localized state exhibits a power-law decay at intermediate timescales dictated by the fractal dimension of the bulk energy states~\cite{Cugliandolo2024} and a metal-insulator transition is reflected in the scattering matrix elements~\cite{MartinezMendoza2013}.

In this work, we perturb the weighted adjacency matrix of the ERG graph by an on-site disorder and the resulting random matrices belong to the sparse Gaussian orthogonal ensemble (sGOE). Such a construction is motivated by the realistic disordered many-body systems, where the presence of impurities leads to diagonal disorder in the matrix representation. While the diagonal disorder lifts the degeneracy of the zero modes present in the adjacency matrix of the ERG graph, we find that the Anderson delocalization-localization transition still occurs at the critical percolation limit $p = N^{-1}$.

We numerically obtain the ground state using the wall-Chebyshev projector and look at the corresponding statistical properties. We find that the distribution of the ground state energy is well described by the Gumbel distribution for $p < N^{-1}$ whereas the Tracy-Widom distribution is valid only at the GOE limit ($p = 1$). The system size scaling of the moments of the ground state intensities indicate that the ground state undergoes a delocalization-localization transition at $\gamma = \gamma_0 \approx 1.86$ where we define $p \equiv N^{-\frac{\gamma}{2}}$.

We analytically compute the energy moments of sGOE and find that the DOS has a transition from the Wigner's semicircle law to Gaussian distribution at $p = N^{-1}$. Later we look at the energy correlations in the bulk spectrum and find that the short-range correlation is present for $p > N^{-1}$  and absent for  $p < N^{-1}$. In the delocalized regime ($p > N^{-1}$), we identify the mobility edge by looking at the ratio of level spacing as a function of energy. The existence of a mobility edge is consistent with the fact the delocalization-localization transition in the ground state is distinct from that in the bulk of the spectrum.

The measures of long-range energy correlations, e.g.~the number variance and the power-spectrum of noise indicate the presence of a Thouless energy scale. Thus, away from the GOE limit  and below the Anderson transition point, we find a nonergodic regime where a localized excitation may take a long time period (inversely proportional to the Thouless energy) to spread over the system compared to a typical ergodic system.

The organization of this paper is as follows: in Sec.~\ref{sec_model}, we introduce the matrix model and its density in the matrix space. 
In Sec.~\ref{sec_ground}, we look at the density of the ground state energy, corresponding fractal dimensions and the entanglement entropy. In Sec.~\ref{sec_DOS}, we show the energy moments and the DOS. In Sec.~\ref{sec_corr}, we show the short- and long-range energy correlations while identifying the mobility edge and Thouless energy scale. In Appendix~\ref{sec_apnd_E_mom}, we show the analytical derivation of the energy moments of our model. In Appendix~\ref{app:ERG_stat}, we discuss the statistical properties of ERG graph. Our concluding remarks are given in Sec.~\ref{sec_Discussion}.

\section{Model}\label{sec_model}
Consider $\mathcal{A}\del{N, p}$, the adjacency matrix of the ERG graph with $N$ number of nodes where $\mathcal{A}_{ii} = 0$ (i.e.~no self-loop) and $\mathcal{A}_{ij} = \mathcal{A}_{ji}$ follows the Bernoulli distribution with a success rate $p$. We analytically compute the energy moments of $\mathcal{A}\del{N, p}$ in Appendix~\ref{ERG_gs}. If we define $p \equiv \frac{b\ln N}{N}$, $\mathcal{A}' \equiv \frac{\mathcal{A}}{\sqrt{b\ln N}}$ and exclude the Perron-Frobenius (largest) eigenvalue of $\mathcal{A}$, the energy spectrum of $\mathcal{A}'$ is within $\pm 2$ for $b > b_\star \equiv (\ln 4 - 1)^{-1}$~\cite{Alt2021a} and all the energy eigenstates are ergodic~\cite{Erdos2012, Erdos2013, He2019}. For $b < b_\star$, mobility edges exist at $E_\star = \pm 2$, separating ergodic states at high energy from the nonergodic extended (NEE) states at low energy~\cite{Alt2021}.

For a NEE state of $\mathcal{A}'$, there are extensive but vanishing fraction of resonant nodes, around which the eigenstate profile has exponential decay within disjoint balls. Corresponding eigenvalues are given by the nodes with abnormally large degree $\geq 2b \ln N$~\cite{Tikhomirov2021}. The system size scaling of the maximum intensity of the energy states shows a sharp discontinuity across the mobility edges. Above properties remain valid for $\frac{\sqrt{\ln N}}{N} \ll p \ll \frac{\ln N}{N}$ whereas localized energy states appear corresponding to the leaves at the periphery of the graph for $p < \frac{\sqrt{\ln N}}{N}$.

We are interested in the spectral properties of the weighted adjacency matrices of sparse graphs. So, we look at the sparse Gaussian orthogonal ensemble (sGOE) consisting of $N \times N$ real symmetric random matrices $H$ with sparsity $0\leq p \leq 1$
\begin{align}
    H = H^\mathrm{Poisson} + H^\mathrm{GOE} \odot \mathcal{A}(N, p)
    \label{eq:sRM_def}
\end{align}
where $H^\mathrm{Poisson}$ is a diagonal matrix from the Poisson ensemble such that $H^\mathrm{Poisson}_{ii} \sim \mathcal{N}\del{0, 1}$, $H^\mathrm{GOE}$ belongs to the Gaussian Orthogonal Ensemble (GOE) such that $H^\mathrm{GOE}_{ij} \sim \mathcal{N}\del{0, \dfrac{1+\delta_{ij}}{2}}$, $\mathcal{N}\del{\bar{x}, \sigma^2}$ is the Gaussian distribution with mean $\bar{x}$ and variance $\sigma^2$ and $A\odot B$ implies the Hadamard product of two matrices $A$ and $B$. The Poisson term in Eq.~\eqref{eq:sRM_def} is motivated by disordered many-body systems where quenched disorder in the on-site interactions adds randomness in the diagonal terms of the matrix representation.

On average, the matrix $H$ from Eq.~\eqref{eq:sRM_def} has $p\binom{N}{2}\sim \mathcal{O}(pN^2)$ off-diagonal elements. Consequently, for $p = 1$ ($p = 0$), we obtain a random matrix from GOE (Poisson ensemble). Note that the DOS of the weighted adjacency matrix of the ERG graph has a divergence at zero energy~\cite{Cugliandolo2024}, see upper panel of Fig.~\ref{fig:ERG_DOS}. Addition of $H^\mathrm{Poisson}$ lifts the degeneracy at zero energy in case of sGOE, see Fig.~\ref{fig:E_mom_sRM}(b).

The matrix model sGOE has important connections to the Sachdev-Ye-Kitaev (SYK) model. In particular, the sparse quadratic SYK model with $N$ Majorana fermions, $\hat{\psi}_i$ is defined as~\cite{GarciaGarcia2024, Orman2025, Granet2025}
\begin{align}
	\hat{H}_{\mathrm{SYK}_2} = \sum_{i < j} J_{ij} p_{ij} \hat{\psi}_i \hat{\psi}_j
\end{align}
where $J_{ij}$ are independent Gaussian random variables and $p_{ij}$ is one (zero) with probability $p$ ($1-p$). In such a sparse SYK model, the dynamics of the real part of the survival amplitude of an initial state can be experimentally simulated on an ion trap quantum computer~\cite{Granet2025}. Note that while the quartic SYK model~\cite{Creffield2026} has been extensively studied due to being the simplest interacting model with a gravity dual, the quadratic SYK model is also important for being an exactly solvable free fermion model~\cite{Winer2020, Lydzba2021, Liu2018}. The single-particle sector of the above Hamiltonian is
\begin{align}
	\hat{H} = \sum_{i < j} J_{ij} p_{ij} \hat{c}_i^\dagger c_j
\end{align}
where $\hat{c}_i^\dagger$ ($\hat{c}_i$) creates (annihilates) a particle at site $i$. As $p_{ij}$ is the $(i, j)$th element of $\mathcal{A}(N, p)$, the adjacency matrix of the ERG graph and $J_{ij}$ (other than a global prefactor) is the $(i, j)$th element of $H^\mathrm{GOE}$, the single-particle representation of the sparse quadratic SYK model is
\begin{align}
	H = H^\mathrm{GOE} \odot \mathcal{A}(N, p).
\end{align}
Then, adding on-site disorder leads to sGOE, the matrix model defined in Eq.~\eqref{eq:sRM_def}. In general, a random matrix from sGOE can be understood as a single-particle model with long-range hopping over 1D lattice with quenched disorder in both on-site potential and hopping terms.

Equation~\eqref{eq:sRM_def} implies that the elements of $H$ have the following distribution
\begin{align}
\begin{split}
    \prob{H_{ij}} &= \begin{cases}
        \dfrac{1}{\sqrt{2\pi}}e^{-\frac{H_{ii}^2}{2}}, & i = j\\
        (1-p)\delta(H_{ij})+\dfrac{p}{\sqrt{\pi}}e^{-H_{ij}^2}, & i\neq j
    \end{cases}
\end{split}
\label{eq:P_Hij_sRM}
\end{align}
The matrix elements, $H_{ij}$'s are independent up to symmetry, leading to a probability density in the matrix space
\begin{align}
\begin{split}
    \prob{H} &= \prod_{i=1}^N \prob{H_{ii}} \prod_{i<j}^{N} \prob{H_{ij}}\\ 
    &= \frac{(1-p)^\frac{N(N-1)}{2}}{(2\pi)^\frac{N}{2}} e^{- \frac{1}{2}\sum\limits_{i = 1}^{N} H_{ii}^2} \prod_{i<j}^{N} \qty[ \delta(H_{ij}) + \frac{p e^{-H_{ij}^2}}{(1-p)\sqrt{\pi}} ]
\end{split}
\label{eq:P_H_sRM}
\end{align}

We want to interpolate between the GOE and Poisson limit by varying the sparsity of sGOE and study corresponding statistical properties. As the critical percolation limit is $p = N^{-1}$, a convenient reparameterization of the sparsity is
\begin{align}
    p \equiv N^{-\frac{\gamma}{2}} { \Rightarrow \gamma = -\frac{2\ln p}{\ln N}}
    \label{eq:g_def}
\end{align}
such that $\gamma = 0$ ($\gamma \to \infty$) is the GOE (Poisson) limit. As the critical percolation limit of ERG graph is $p = N^{-1}$, $\gamma = 2$ should be the critical point where delocalization-localization transition occurs, which we establish later via energy correlations.

\section{Ground state}\label{sec_ground}

 Given an $N \times N$ matrix, we denote its sorted energy levels (in ascending order) as $\{ E_1, E_2,\dots, E_N \}$ such that $E_1$ is the energy of the ground state. In case of the Poisson ensemble, the energy levels follow the Gaussian distribution with a tail decaying faster than the power-law while the domain is unbounded. Consequently, the ground state energy of the Poisson ensemble follows the reflected Gumbel distribution~\cite{Haan2006book}
\begin{align}
    \mathrm{P}_{\text{E}_1}(y) = \frac{1}{\sEV} \exp\left( \frac{y + \lEV}{\sEV} - \exp\left(\frac{y + \lEV}{\sEV} \right) \right).
    \label{eq:P_E1_gumbel}
\end{align}
where the location ($\lEV$) and scale ($\sEV$) parameters are~\cite{Kotz2000book}
\begin{align}
    \lEV = g\left (1 - \frac{1}{N}\right ), \quad \sEV = g \left (1 - \frac{1}{Ne} \right ) - \lEV
\end{align}
where $g(x) = \sqrt{2}\sgE\mathrm{erf}^{-1}(2x - 1)$ is the inverse of the cumulative distribution of energy for the Poisson ensemble.

\begin{figure}[t]
    \centering
    \includegraphics[width=0.9\columnwidth]{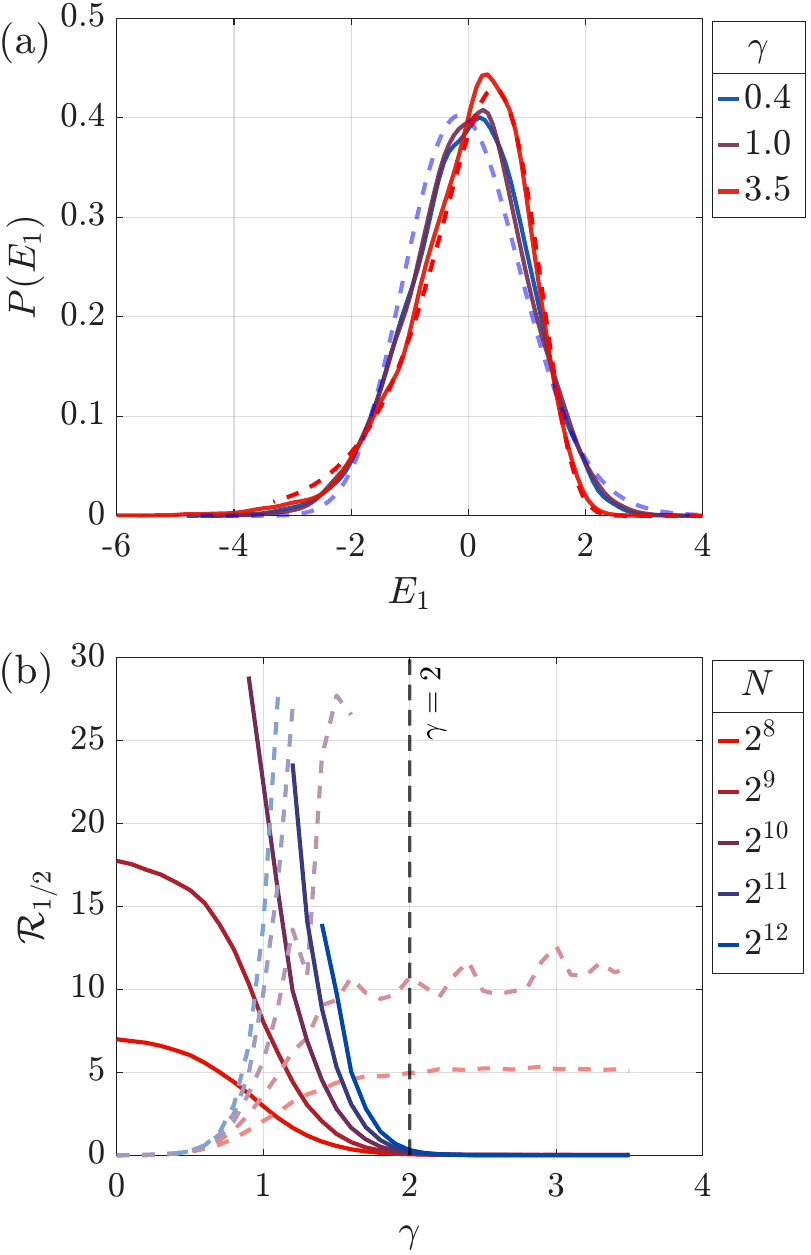}
    \caption[Density of ground state for sGOE]{{\bfseries Ground state energies of sGOE.} (a)~density of the ground state energies obtained using the wall-Chebyshev projector~\cite{Zhao2024} for $N = 16384$ and different values of $\gamma$. The energy levels are shifted and scaled to ensure zero mean and unit variance. Dashed red (blue) curve denotes the Gumbel (Tracy-Widom) distribution.
    (b)~R\'enyi divergence of order $q = \frac{1}{2}$ between the numerical histogram of the ground state and the Gumbel (Tracy-Widom) distribution as a function of $\gamma$ for different system sizes shown via solid (dashed) lines.
    }
    \label{fig:DOGS}
\end{figure}

In the case of Wigner-Dyson ensembles, the extreme eigenvalues follow the Tracy-Widom distributions~\cite{Tracy1994, Tracy1996, Nadal2011}. The average ground state energy for the GOE is~\cite{Johnstone2012}
\begin{align}
    \begin{split}
        \mean{E_1} &= -\frac{\sgE}{\sqrt{\frac{N+1}{2}}}\del{\lEV + \sEV E_\mathrm{TW}^{(1)}}\\
        \lEV^2 &= 2\del{N - \frac{1}{2} - \frac{(N - \frac{1}{2})^{-\frac{1}{3}}}{10}},\: \sEV = \frac{N^{-\frac{1}{6}}}{\sqrt{2}}
    \end{split}
    \label{eq:TW_params}
\end{align}
where $E_\mathrm{TW}^{(1)} = -1.20653357458\dots$ is the expectation value w.r.t.~Tracy-Widom distribution with unit Dyson index.

In the case of sGOE, we expect the Gumbel  distribution to be valid in the limit $p\to 0$  and Tracy-Widom for $p\to 1$. To study the distribution of the lowest energy of sGOE for intermediate values of $p$, we obtain the corresponding ground state using the wall-Chebyshev projector~\cite{Zhao2024}. Such an algorithm is based on the fact that the wall function
\begin{align}
    \mathrm{wall}(x) = \begin{cases}
        \infty, & x < 0\\
        1, & x = 0\\
        0, & x > 0
    \end{cases}
\end{align}
can be expanded in terms of the Chebyshev polynomials and $\mathrm{wall}(H)$ applied to a random vector converges to the ground state of the Hamiltonian $H$.

In Fig.~\ref{fig:DOGS}(a), we show the density of ground state energy for different values of $\gamma$. To understand the validity of the Gumbel or Tracy-Widom distribution at different values of the sparsity parameter, we compute the R\'enyi divergence of order $q$~\cite{Nagy2015}
\begin{align}
    \mathcal{R}_q = \frac{1}{q - 1}\ln\del{\int dx \del{\mathrm{P}(x)}^q \del{\mathrm{P}_\mathrm{test}(x)}^{1-q} }
\end{align}
where $\mathrm{P}(x)$ is the numerically obtained histogram of the ground state energy and $\mathrm{P}_\mathrm{test}(x)$ is either Gumbel or Tracy-Widom distribution. 
In Fig.~\ref{fig:DOGS}(b), we show the R\'enyi divergences as a function of $\gamma$ for different system sizes. Our analyses show that if the test function is Gumbel distribution, the R\'enyi divergence increases (decreases) with system size for $\gamma < 2$ ($\gamma > 2$), i.e.~the ground state energy of sGOE follows the Gumbel distribution for $\gamma > 2$, provided $N \gg 1$.\linebreak
This is expected as $\gamma=2$ is the Anderson delocalization-localization transition point, beyond which the energy states are localized with uncorrelated energy levels. On the other hand, the R\'enyi divergence w.r.t.~Tracy-Widom distribution is zero only at GOE limit, $\gamma = 0$, and increases with $N$ for $\gamma > 0$.

\subsection*{Structure of ground state}

\begin{figure}[t]
    \centering
    \includegraphics[width=0.9\columnwidth]{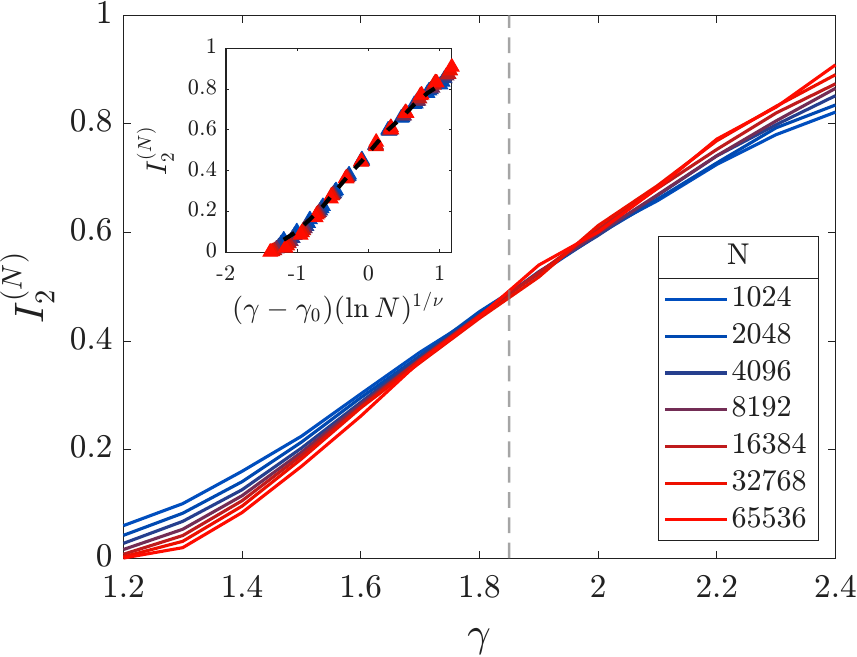}
    \caption[Ground state IPR]{ Inverse participation ratio (Eq.~\eqref{eq:def_Iq}) of the ground state of sGOE for $q = 2$ as a function of $\gamma$ for different system sizes. Inset shows the collapsed data using the ansatz in Eq.~\eqref{eq:2nd_ansatz} giving $\gamma_0 \approx 1.862$, shown by the vertical dashed line.
    }
    \label{fig:IPR_grnd}
\end{figure}

The structure of the ground state, in particular the localization length and the fluctuations of the ground state intensities determine the nature of dynamics at low-temperature for an isolated quantum system. To understand the localization properties of the ground state over the Hilbert space, we look at the generalized inverse participation ratios (IPRs)~\cite{Soukoulis1984, Rodriguez2010, Evers2008, Das2024, Das2025b}
\begin{align}
    I_q^{(N)} = \sum_{n = 1}^N |\Psi(n)|^{2q} \sim N^{(1-q)D_q^{(N)}}
    \label{eq:def_Iq}
\end{align}
where $\Psi(n) = \bra{\Phi_n}\ket{\Psi}$ is the $n$th component of the state vector $\ket{\Psi}$ in a chosen basis $\cbr{\ket{\Phi_n}}$, $N$ is the Hilbert space dimension and $D_q^{(N)}$ is the $q$th finite size fractal dimension. In Fig.~\ref{fig:IPR_grnd}, we show the ensemble averaged IPR ($q=2$) of the ground state of sGOE and different system sizes ranging from $N = 1024$ to $65536$. The inset of Fig.~\ref{fig:IPR_grnd} shows that the IPR data for different values of $N$ can be collapsed using the 2nd order transition ansatz~\cite{Pino2019, Das2022, Das2022b, Das2026}
\begin{align}
    \ot(\gamma, N) = f((\gamma-\gamma_0)[\ln{N}]^\frac{1}{\nu})
    \label{eq:2nd_ansatz}
\end{align}
where $\ot$ is the observable of our interest, $f$ is an universal function, $\gamma_0$ and $\nu$ are the critical parameter and exponents, respectively. In particular, we find that $\gamma_0 \approx 1.862$ for the IPR of the ground state of sGOE, implying a delocalization-localization transition at zero temperature. 
In Fig.~\ref{fig:sGOE_ERG_compare}, we compare the IPR of the ground states of ERG and sGOE for different system sizes. Importantly, the ground state phase transition occurs at $\gamma_0 \approx 1.63$ for the adjacency matrix of ERG graph (see Fig.~\ref{fig:ERG_GS}). Thus, addition of quenched disorder in the on-site potential and hopping terms of a sparse graph localizes the ground state at smaller values of sparsity.

To understand the presence of different length scales in the fluctuations of the ground state intensities, we obtain $D_q^{(N)}$, the finite size fractal dimensions from the system size scaling of the IPRs. Then, we extrapolate $D_q^{(N)}$ to the thermodynamic limit (assuming that the finite size corrections are polynomial functions of $1/\ln N$~\cite{Luitz2020}) and extract the fractal dimensions, $D_q$. The fractal dimensions can at most monotonically decrease with $q$ and are bounded from below by $D_\infty$, which determines the scaling of the maximum wavefunction intensity, $\max_n |\Psi(n)|^2 \sim N^{-D_\infty}$~\cite{Lakshminarayan2008, Lindinger2019}. All the fractal dimensions are unity for an ergodic wavefunction, which is uniformly spread over the Hilbert space irrespective of the choice of basis~\cite{Goldstein2010, Vikram2023}. On the other hand, $D_q = 0$ for all values of $q$ for a wavefunction localized on the Fock space.

\begin{figure}[t]
    \centering
    \includegraphics[width=0.9\columnwidth]{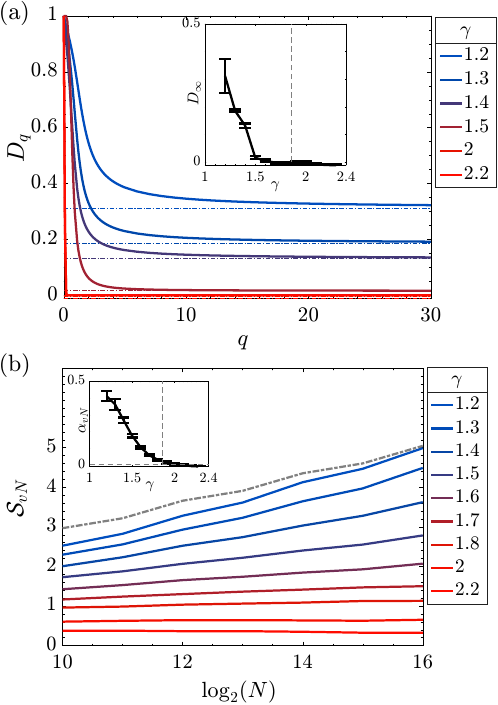}
    \caption[Ground state fractal dimensions and entanglement entropy]{{\bfseries Ground state of sGOE.} (a)~Fractal dimensions of the ground state for different values of $\gamma$. Horizontal dashed lines denote the smallest fractal dimension, $D_\infty$. Inset shows $D_\infty$ as a function of $\gamma$. 
    (b)~Bipartite von Neumann entanglement entropy as a function of system size for various $\gamma$. Gray dashed line denotes the Page value~\cite{Page1993}. Inset shows the system size scaling $\mathcal{S}_{vN} \propto \alpha_{vN} \ln N$. Error-bars denote 95\% confidence interval. Vertical dashed lines in the insets indicate ground state delocalization-localization transition at $\gamma_0 \approx 1.86$}. 
    
    \label{fig:Dq_grnd}
\end{figure}

In Fig.~\ref{fig:Dq_grnd}(a), we show the fractal dimensions for different values of $\gamma$. We find that for $\gamma \geq \gamma_0$, the fractal dimensions are zero for $q>0$, indicating that the ground state is localized. Contrarily for $\gamma < \gamma_0$, $0 < D_q < 1$ and the fractal dimensions exhibit $q$-dependence. This implies that different moments of the ground state intensities exhibit distinct scaling exponents, i.e.~the ground state is multifractal~\cite{Das2025b}. Thus, the ground state is nonergodic but extended for $0 < \gamma < \gamma_0$, with an extensively large support set on the Hilbert space but occupies a vanishing fraction of the Hilbert space volume~\cite{Kravtsov2015, DeLuca2014, Das2022, Roy2023}. In particular, $D_\infty$ (see the inset of Fig.~\ref{fig:Dq_grnd}(a)) indicates that a vanishing fraction of the ground state intensities are anomalously large in the delocalized regime.

To understand the structure of an eigenstate in case of a random matrix, the site basis is the natural choice for computing the IPRs. However, the choice of basis can be ambiguous for a generic nonergodic system. Compared to the localization measures like IPRs, the von Neumann entanglement entropy provides a basis independent analysis of a wavefunction~\cite{Popescu2006, Amico2008, Sinha2020}. To compute the entanglement measures in case of a random matrix, consider two Hilbert spaces, $\mathcal{H}_A$ and $\mathcal{H}_B$ with dimensions $N_A$ and $N_B$ and spanned by the bases, $\cbr{\ket{a_i}}$ and $\cbr{\ket{b_j}}$, respectively. Then, any pure state $\ket{\Psi}$ in the tensor product space $\mathcal{H}\equiv \mathcal{H}_A \otimes \mathcal{H}_B$ can be expressed as~\cite{DeTomasi2020a, Haque2022}
\begin{align}
	\label{eq_V_tensor}
	\ket{\Psi} = \sum_{i, j} c_{ij} \ket{a_i} \otimes \ket{b_j}
\end{align}
where $\cbr{c_{ij}}$ is a $N_A\times N_B$ matrix. Schmidt decomposition implies that $\cbr{c_{ij}} = U\Sigma V^\ast$ ($= U\Sigma V^T$) where $U, V$ are $N_A\times N_A$ and $N_B\times N_B$ unitary (orthogonal) matrices for complex (real) matrix $\cbr{c_{ij}}$. $\Sigma$ is a $N_A\times N_B$ diagonal matrix with real non-negative entries, $\cbr{\alpha_k}_{1\leq k\leq N'}$, where $N'\leq \min\cbr{N_A, N_B}$ is the rank of $\cbr{c_{ij}}$. The Schmidt coefficients, $\alpha_k$'s are unique up to reordering but $U, V$ are not. Then,
\begin{align}
	\ket{\Psi} = \sum_{k = 1}^{N'} \alpha_k \ket{u_k} \otimes \ket{v_k}
\end{align}
where $\ket{\Psi}$ is a column vector with elements $c_{ij}$ from Eq.~\eqref{eq_V_tensor} and $\ket{u_k}, \ket{v_k}$ are the column vectors of $U, V$. $\ket{\Psi}$ is a separable state for $N' = 1$ and entangled state for $N'>1$. For a pure state $\varrho\equiv \ket{\Psi}\bra{\Psi}$, partial trace over $\mathcal{H}_A$ or $\mathcal{H}_B$ gives a matrix with $|\alpha_k|^2 \equiv \lambda_k$ as the diagonal elements.

If an $N$-dimensional Hilbert space $\mathcal{H}$ can be decomposed as a tensor product space $\mathcal{H}\equiv \mathcal{H}_A \otimes \mathcal{H}_B$, where $\mathcal{H}_{A, B}$ corresponds to the subsystems $A$ and $B$ with dimensions $N_A$ and $N/N_A$, the reduced density matrix of a pure state $\ket{\Psi}$ in $\mathcal{H}$ over subsystem $A$ is $\varrho_A \equiv \Tr_B\sbr{\ket{\Psi}\bra{\Psi}}$. Then, the von-Neumann entanglement entropy of $\ket{\Psi}$ w.r.t.~the subsystem $A$ is~\cite{Buijsman2026}
\begin{align}
	\label{eq:EE_def}
	\mathcal{S}_{vN} = -\sum_{k = 1}^{N'} \lambda_k \ln \lambda_k,\quad N' \leq \min\cbr{N_A, \frac{N}{N_A}}
\end{align}
where the diagonal elements of $\varrho_A$ are $\lambda_k$. Since $\sum_{k} \lambda_k = 1$ and $0 < \lambda_k \leq 1$, $\lambda_k$'s are distributed over $(N'-1)$-dim standard simplex. The eigenvalues of $-\ln \varrho_A$, determines the entanglement spectrum, $\cbr{-\ln \lambda_k}$~\cite{Sinha2021}.

The entanglement entropy, $\mathcal{S}_{vN}$ quantifies the degree of correlation between the subsystems $A$ and $B$ over a given state. $\mathcal{S}_{vN}$ is 0 for separable states whereas for Haar random vectors, any two subsystems are maximally entangled and the entanglement entropy follows the Page value~\cite{Page1993}
\begin{align}
	\mathcal{S}_{vN} \approx \ln N_A - \frac{N_A^2}{2N},\qquad N_A^2\leq N.
	\label{eq:EE_Haar}
\end{align}
The maximal entanglement entropy from Eq.~\eqref{eq:EE_Haar} as a function of the subsystem size is known as the Page curve. In Fig.~\ref{fig:Dq_grnd}(b), we show the bipartite von Neumann entanglement entropy of the ground state of sGOE as a function of system size for different values of $\gamma$. We find that the entanglement entropy is independent of system size for $\gamma \geq \gamma_0$ while a linear growth $\mathcal{S}_{vN} \sim \alpha_{vN} \log_2 N$ is observed for $\gamma < \gamma_0$. The prefactor $\alpha_{vN}$ is shown in the inset of Fig.~\ref{fig:Dq_grnd}(b) as a function of $\gamma$. We find that $0 < \alpha_{vN} < \frac{1}{2}$ for $0 < \gamma < \gamma_0$ whereas $\alpha_{vN} \approx \frac{1}{2}$ for an ergodic state. Thus, the entanglement entropy analysis is consistent with our localization measures: the ground state is localized for $\gamma > \gamma_0$ and nonergodic extended for $0 < \gamma < \gamma_0$, yielding a zero-temperature delocalization-localization transition at $\gamma_0 \approx 1.86$. On the contrary, similar transition occurs at $\gamma_0 \approx 1.63$ for the adjacency matrix of the ERG graph as shown in Appendix~\ref{app:ERG_stat}. In Fig.~\ref{fig:sGOE_ERG_compare}, we compare the entanglement entropy of the ground states of ERG and sGOE for different system sizes.

The existence of the above critical point can be argued as follows. For $\frac{1}{N} \lessapprox p \lessapprox \frac{\ln N}{N}$, the ERG graph has a giant component having $\mathcal{O}(sN)$ vertices with average degree $\sim Np(2 - s)$ where $s$ follows the self-consistent equation $s = 1 - e^{-pN s}$~\cite{Durrett2007book}. If we take $pN = 1 + \epsilon$ with $\epsilon \ll 1$, the giant connected component has $\mathcal{O}(2\epsilon N)$ number of vertices with average degree $Np(2 - 2\epsilon) \approx 2$. In a tree-like graph with degree $k$, Anderson transition occurs at the critical disorder strength $\mathcal{O}(k\ln k)$~\cite{AbouChacra1973, AbouChacra1974} whereas all the eigenstates are extended for disorder strength smaller than $\mathcal{O}(k)$~\cite{Aizenman2011}. 

As the giant component is an approximate tree-like graph with vertex degree 2 for $pN = 1+\epsilon$, the critical disorder strength should be 2 below which the ground state cannot localize whereas we sample the on-site disordered potentials from Gaussian distribution with unit variance. The ground state of sGOE should localize only when the graph gets fragmented with a global bandwidth $\sim\mathcal{O}(1)$ for $pN < 1$. Note that for a finite system size $N$, the smallest on-site disorder follows the Gumbel distribution such that we cannot rule out a logarithmic correction to $p = \frac{1}{N}$.Numerically we determine the ground state delocalization-localization transition to be $\gamma_0 \approx 1.86$ which is lower than the Anderson transition in the bulk expected at $\gamma = 2$. Since for a range of $\gamma$ values, the ground state localizes while the bulk eigenstates are still delocalized, we expect the presence of a mobility edge in case of sGOE. In Sec.~\ref{sec_corr}, we confirm this by looking at energy resolved short-range correlations.

\section{Density of states}\label{sec_DOS}
In this section, we show the ensemble averaged energy moments whose details are given in Appendix~\ref{sec_apnd_E_mom}.
Given the density in the matrix space (Eq.~\eqref{eq:P_H_sRM}) the ensemble average of the second moment of the energy
\begin{align}
    \expval{\expval{E^2}} = \frac{1}{N}\qty(N + \frac{N(N-1)p}{2}) = 1+\frac{p}{2}(N-1)
    \label{eq:E2_sRM_m}
\end{align}
It is also possible to estimate the ensemble averaged fourth moment of energy for sGOE
\begin{align}
    \mean{\expval{E^4}}
    = \frac{p^2 N^2}{2}
    + \left(\frac{11}{4}p - \frac{3}{2}p^2\right) N
    + \left(p^2 - \frac{11}{4}p + 3\right)
    \label{eq:E4_sRM_m}
\end{align}
The analytical expressions for the energy moments of the ERG graph are given in Appendix~\ref{ERG_gs}.

Above equations imply that for $p = N^{-1}$, $\mean{\mean{E^2}} \approx \dfrac{3}{2}$ and $\mean{\mean{E^4}} \approx \dfrac{25}{4}$, provided $N\gg 1$. We numerically obtain the energy moments using the Girard-Hutchinson estimator~\cite{Epperly2024}
\begin{align}
    \Tr H^n \approx \frac{1}{m}\sum_{j = 1}^{m} \bra{\Psi_j} H^n \ket{\Psi_j}
    \label{eq:GH_estimator_def}
\end{align}
where $\ket{\Psi_j}$'s are normalized random vectors and $1\ll m\ll N$. The above stochastic sampling allows us to obtain the energy moments without doing exact diagonalization of $H$. We checked that the numerically obtained energy moments have a nice agreement with our analytical expressions.

\begin{figure}[t]
    \includegraphics[width=\linewidth]{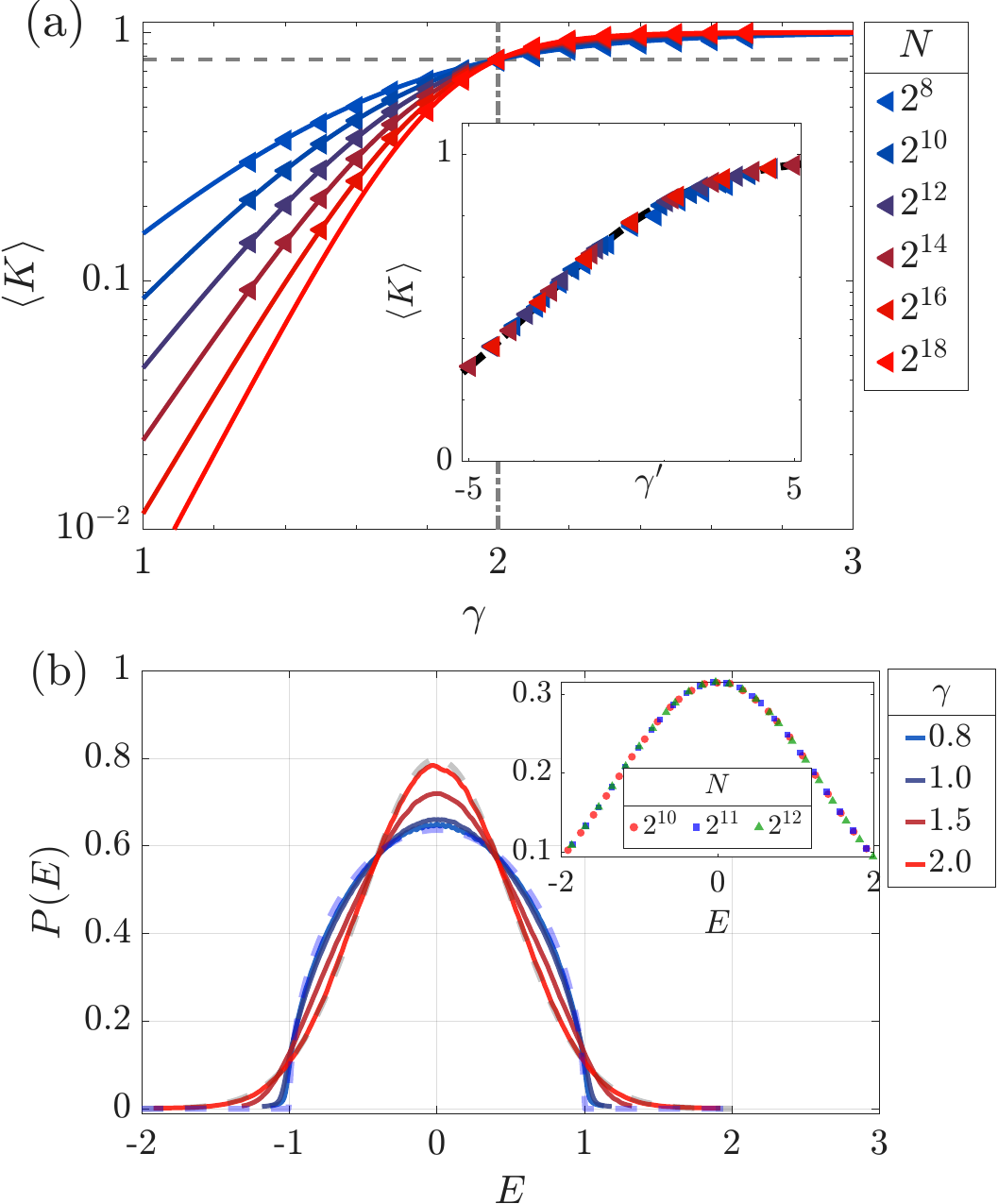}
    \caption{{\bfseries Energy moments and DOS of sGOE.} (a)~shifted kurtosis (Eq.~\eqref{eq:Kurt_def}) as a function of $\gamma$ for different system sizes (values given in the legend) where sparsity $p \equiv N^{-\frac{\gamma}{2}}$. Markers denote numerical data obtained using Girard-Hutchinson estimator (Eq.~\eqref{eq:GH_estimator_def}) and the solid lines denote the analytical expressions from Eq.~\eqref{eq:kurt}. Horizontal dashed line correspond to $\mean{K} = \frac{25}{4}$. The inset shows the collapsed data assuming 2nd order transition ansatz (Eq.~\eqref{eq:2nd_ansatz}) where $\gamma' \equiv (\gamma - 2)\ln N ^{\frac{1}{\nu}}$, $\nu \approx 1.0464$.
    (b)~ensemble averaged DOS obtained using the Lanczos method~\cite{Lin2016} for $N = 1024$ and different values of $\gamma$. We fix the energy variance $\sgE^2 = \frac{1}{4}$ for all values of $\gamma$. Dashed red (blue) curve denotes the Gaussian (semicircle) distribution. Inset shows the average DOS at $\gamma = 2$ for different system sizes.
    }
    \label{fig:E_mom_sRM}
\end{figure}

Based on the 2nd and 4th moments of energy, we define the \textit{shifted kurtosis}
\begin{align}
    K = \frac{\expval{E^4}}{\expval{E^2}^2} - 2.
    \label{eq:Kurt_def}
\end{align}
The kurtosis allows us to quantify the change in the shape of the density of states (DOS)
\begin{align}
    \rho(E) = \frac{1}{N} \sum_{i=1}^N \delta(E-E_i)
    \label{eq:DOS_def}
\end{align}
where $\cbr{E_j}$ are the $N$ number of energy levels of a system. The DOS measures the number of available energy eigenstates in a given energy window. In case of a Gaussian (semicircle) DOS, the kurtosis has a value $K = 1$ ($K = 0$). Thus, we expect the kurtosis to interpolate between 0 and 1 as we increase the sparsity of a random matrix from sGOE. The analytical expressions of the energy moments (Eqs.~\eqref{eq:E2_sRM_m} and \eqref{eq:E4_sRM_m}) give us
\begin{equation}\label{eq:kurt}
    K = \frac{4-p(N-1)(2p-3)}{(2+(N-1)p)^2}
\end{equation}
For $p = N^{-1}$, $K = \dfrac{7}{9}$ is independent of the system size in the limit $N\to\infty$. In Fig.~\ref{fig:E_mom_sRM}(a), we show the numerically obtained kurtosis along with the analytical expression in Eq.~\eqref{eq:kurt} as a function of $\gamma$ and find a good agreement. In the inset of Fig.~\ref{fig:E_mom_sRM}(a), we show that the kurtosis data from different system sizes can be collapsed assuming the ansatz in Eq.~\eqref{eq:2nd_ansatz}.

For $N\gg 1$, the shifted kurtosis becomes,
\begin{equation}
    K = \frac{4-pN(2p-3)}{\qty(2+pN)^2}.
\end{equation}
Defining the variable $x = (\gamma -2) (\ln N)$ we have,
\begin{equation}
    pN = N^{1-\gamma/2} = \exp \qty[-\frac{1}{2}(\gamma -2)\ln N] \equiv e^{-x/2}
\end{equation}
For $N\gg 1$, we have $p = N^{-\gamma/2} \to 0$ using which we get,
\begin{align}
    \begin{split}
        K &= \frac{4 - e^{-x/2}(2p-3)}{\qty(2+e^{-x/2})^2} = \frac{4 +3 e^{-x/2}}{\qty(2+e^{-x/2})^2}\\
        &= f\qty( (\gamma -2) (\ln N))
    \end{split}
\end{align}
from which we get $\nu=1$ in close agreement with our numerical collapse in Fig.~\ref{fig:E_mom_sRM} .A fundamentally different behavior for the shifted kurtosis is seen for the adjacency matrix of the ERG graph, shown in Fig.~\ref{fig:ERGkurt} of Appendix~\ref{app:ERG_stat} with analytical expression given in Eq.~\eqref{app:en_kurt}.
 
The behavior of Kurtosis indicate that the DOS of sGOE should be independent of the system size at $\gamma = 2$ (i.e.~$p = N^{-1}$), which we verify next.

In the limit $p = 0$, a matrix from sGOE belongs to the Poisson ensemble and the corresponding average DOS follows a Gaussian distribution
\begin{align}
    \mean{\rho^\mathrm{P}(E)} = \frac{1}{\sqrt{2\pi}\sgE}\exp\del{-\frac{E^2}{2\sgE^2}}
    \label{eq:DOS_avg_Psn}
\end{align}
where 
\begin{align}
    \sgE^2 \equiv \mean{E^2} - \mean{E}^2
\end{align}
is the energy variance. On the other hand, for $p = 1$, we obtain a matrix from the GOE, where the average DOS follows the Wigner's semicircle law for $N\to\infty$
\begin{align}
    \rho^\mathrm{SC}(E) = \frac{\sqrt{4\sgE^2 - E^2}}{2\pi \sgE^2}.
    \label{eq:DOS_SC}
\end{align}
For finite $N$, the DOS of GOE can be expressed in terms of the Hermite polynomials~\cite{Livan2018book}, with a deviation at the edge of the spectrum from the above semicircle law. Nevertheless, $\rho^\mathrm{SC}(E)$ well approximates the average DOS in the bulk energy spectrum of the GOE for $N\gg 1$. In case of sGOE, we expect the average DOS to interpolate between the semicircle law (Eq.~\eqref{eq:DOS_SC}) and the Gaussian distribution (Eq.~\eqref{eq:DOS_avg_Psn}) as we increase the sparsity parameter. This is contrary to the ERG graph where the DOS exihibits a peak at $E=0$ on increasing sparsity, due to the absence of diagonal disorder. This has been verified in Fig. \ref{fig:ERG_DOS} of Appendix \ref{app:ERG_stat}.

We numerically estimate the ensemble averaged DOS using the kernel polynomial method (KPM)~\cite{Silver1994, Wang1994, Weisse2006}. First, we estimate the lower and upper energy bounds using a partial Lanczos reduction of the input matrix $H$~\cite{Lin2016}. Then, we scale and shift $H$ to ensure the energy levels are within [-1, 1]. Next, we expand the DOS w.r.t.~the Chebyshev polynomials of 1st kind, $T_j(x)$
\begin{align}
    \rho(\tilde{E}) = \frac{1}{\sqrt{1 - \tilde{E}^2}} \sum_{j = 0}^{\infty} \mu_j T_j(\tilde{E})
    \label{eq:DOS_Tn_expansion}
\end{align}
where $-1\leq \tilde{E} \leq 1$ is the shifted and scaled energy. The Chebyshev moments, $\mu_j$'s are uniquely related to the energy moments due to the orthogonality of the Chebyshev polynomials. For example,
\begin{align}
    \mean{\tilde{E}^2} = \frac{1}{2} + \frac{\pi \mu_2}{4},\quad \mean{\tilde{E}^4} = \frac{3}{8} + \frac{\pi \mu_2}{4} + \frac{\pi \mu_4}{16}
\end{align}
To numerically estimate the DOS, we truncate the series in Eq.~\eqref{eq:DOS_Tn_expansion} to a finite order and estimate the $\mu_j$'s using a stochastic sampling as in Eq.~\eqref{eq:GH_estimator_def}. Such a truncation leads to a Gibbs oscillation in the estimated DOS which is a sum of delta functions. We convolve the DOS with the Jackson kernel to get rid of the Gibbs oscillation~\cite{Silver1994}. We find that despite Jackson damping, the estimated DOS tends to have negative values close to the edges of the spectrum.

To avoid the unphysical situation where the DOS becomes negative, we use the Lanczos method~\cite{Haydock1975, Lin2016} to estimate the DOS. We perform a partial Lanczos reduction of $H$ w.r.t.~a random vector to obtain the Ritz values and vectors. A weighted histogram of the Ritz values w.r.t.~the 1st squared components of the Ritz vectors gives an estimate of the DOS.

In Fig.~\ref{fig:E_mom_sRM}(b), we show the ensemble averaged DOS for different values of $\gamma$ obtained using the Lanczos method. The obtained DOS is always positive in contrast to the KPM. Note that, the DOS can also be obtained by numerically solving the Hammerstein equations, as shown for the adjacency and Laplacian matrices of the ERG graph in Ref.~\cite{Akara-pipattana2025}.

In the inset of Fig.~\ref{fig:E_mom_sRM}(b), we show the average DOS for $\gamma = 2$ and different system sizes. We find that the DOS is independent of $N$, as predicted by the shifted kurtosis in Eq.~\eqref{eq:kurt}. We compared the numerically obtained DOS of sGOE with the analytical DOS of other notable random matrix models, e.g.~Rosenzweig-Porter ensemble~\cite{Bertuola2005}, banded random matrices~\cite{Das2025a}, $\beta$-ensemble~\cite{Roy2025}. We do not find an one-to-one correspondence of the sGOE with any of the above random matrix models in terms of the average DOS, as the crossover from the semicircle to the Gaussian DOS is not universal.

\section{Energy correlations}\label{sec_corr}
The energy levels of a matrix from GOE are correlated at all possible length scales, which is a characteristic signature of quantum chaos~\cite{Bohigas1984, Shklovskii1993}. Contrarily, the energy levels of the Poisson ensemble remain uncorrelated as conjectured  for an integrable quantum system~\cite{Berry1977}. In case of sGOE, we intuitively expect that the energy correlations get suppressed as we increase the sparsity of the random matrix. To quantify the nature of short-range energy correlation (i.e.~on the scale of mean level spacing), we look at the distribution of level spacings, $s_j \equiv \Ecal_{j+1} - \Ecal_{j}$, where $\cbr{\Ecal_j}$ is the unfolded spectrum~\cite{Mehta2004book, Guhr1998}. For unfolded spectrum, the DOS becomes uniform such that the mean position of the $j$th unfolded energy level $\mean{\Ecal_j} = j$. 
In case of  the Poisson ensemble, the density of level spacings follows an exponential distribution, $\pPsn{s} = e^{-s}$ while for GOE, the density of level spacing is well approximated by the Wigner's surmise~\cite{Dietz1990}. The Wigner's surmise implies that $\pGOE{s \to 0} \to 0$, i.e.~the energy levels cannot come arbitrarily close to each other, a property known as the \textit{level repulsion}, which is typically observed in chaotic quantum systems~\cite{Bohigas1984, Shklovskii1993, Das2019, Das2023a}.

\begin{figure}[t]
    \includegraphics[width=0.8\linewidth]{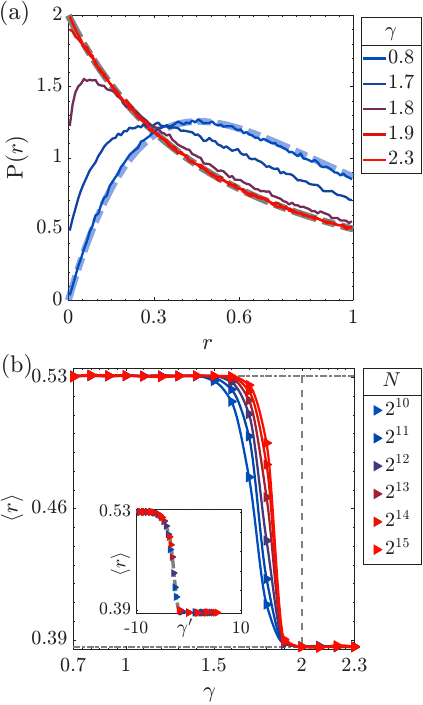}
    \caption[Short-range energy correlation of sGOE]{{\bfseries Short-range energy correlation of sGOE.} (a)~density of the ratio of level spacings for middle 40\% spectrum, $N = 1024$ and different values of $\gamma$. Dashed blue (red) curve denotes the analytical expressions of GOE (Poisson ensemble). (b)~average ratio of level spacings as a function of $\gamma$ for different system sizes. Inset shows the collapsed data following Eq.~\eqref{eq:2nd_ansatz} for a critical parameter $\gamma_c = 2$ and exponent $\nu \approx 0.9316$.}
    \label{fig:short-range_sRM}
\end{figure}

We find that for small values of sparsity $\gamma$, the density of level spacings $\prob{s}$ follows that of the GOE, while upon increasing $\gamma$ the density converges to that of the Poisson ensemble~\cite{Jackson2001}. To quantify such a crossover and ensuring that our analysis of the short-range correlation is not affected by the unfolding process~\cite{Gomez2002}, we look at the ratio of level spacings~\cite{Oganesyan2007, Atas2013}
\begin{align}
    r_j = \min\del{\tilde{r}_j, \frac{1}{\tilde{r}_j}},\quad \tilde{r}_j \equiv \frac{E_{j+1} - E_j}{E_j - E_{j-1}}.
\end{align}
In case of the GOE and the Poisson ensemble, the density of the ratio of level spacings follows~\cite{Atas2013}
\begin{align}
    \begin{split}
        \pGOE{r} &= \frac{27}{4}\frac{r+r^2}{(1+r+r^2)^\frac{5}{2}}\\
        \pPsn{r} &= \frac{2}{(1+r)^2},\quad 0\leq r \leq 1.
    \end{split}
    \label{eq:r_GOE_Psn}
\end{align}
In Fig.~\ref{fig:short-range_sRM}(a), we show the density of the ratio of level spacings for the middle 40\% energy spectrum for $N = 1024$ and different values of $\gamma$ along with the above analytical expressions. We find that  $\prob{r}$ interpolates between the GOE and Poisson limit as we increase $\gamma$ and to quantify such interpolation, we look at the average ratio of level spacing, $\mean{r}$. From the analytical expressions in Eq.~\eqref{eq:r_GOE_Psn}, we can calculate the average ratio of level spacing to be $\mean{r}^\mathrm{GOE} \approx 0.53$ and $\mean{r}^\mathrm{Poisson} \approx 0.39$ for the GOE and Poisson ensemble, respectively~\cite{Atas2013}.

In Fig.~\ref{fig:short-range_sRM}(b), we show the average ratio of level spacing as a function of $\gamma$ for different system sizes. Corresponding inset shows the collapsed data following Eq.~\eqref{eq:2nd_ansatz}. Thus, both the level spacing and its ratio indicate that the short-range energy correlation in the bulk spectrum of sGOE undergoes a 2nd order phase transition at $\gamma = 2$ (i.e.~sparsity, $p = N^{-1}$) such that there is no correlation among the neighboring bulk energy levels for $\gamma > 2$. A similar scaling is observed for the ERG graph after removing the zero modes, as shown in Fig.~\ref{fig:ERG_DOS} of Appendix~\ref{app:ERG_stat}. Thus, short-range energy correlations show that the Anderson transition in case of sGOE occurs at the critical percolation limit, $\gamma = 2$.

\subsection{Mobility edge}
\begin{figure}[t]
    \includegraphics[width=\linewidth]{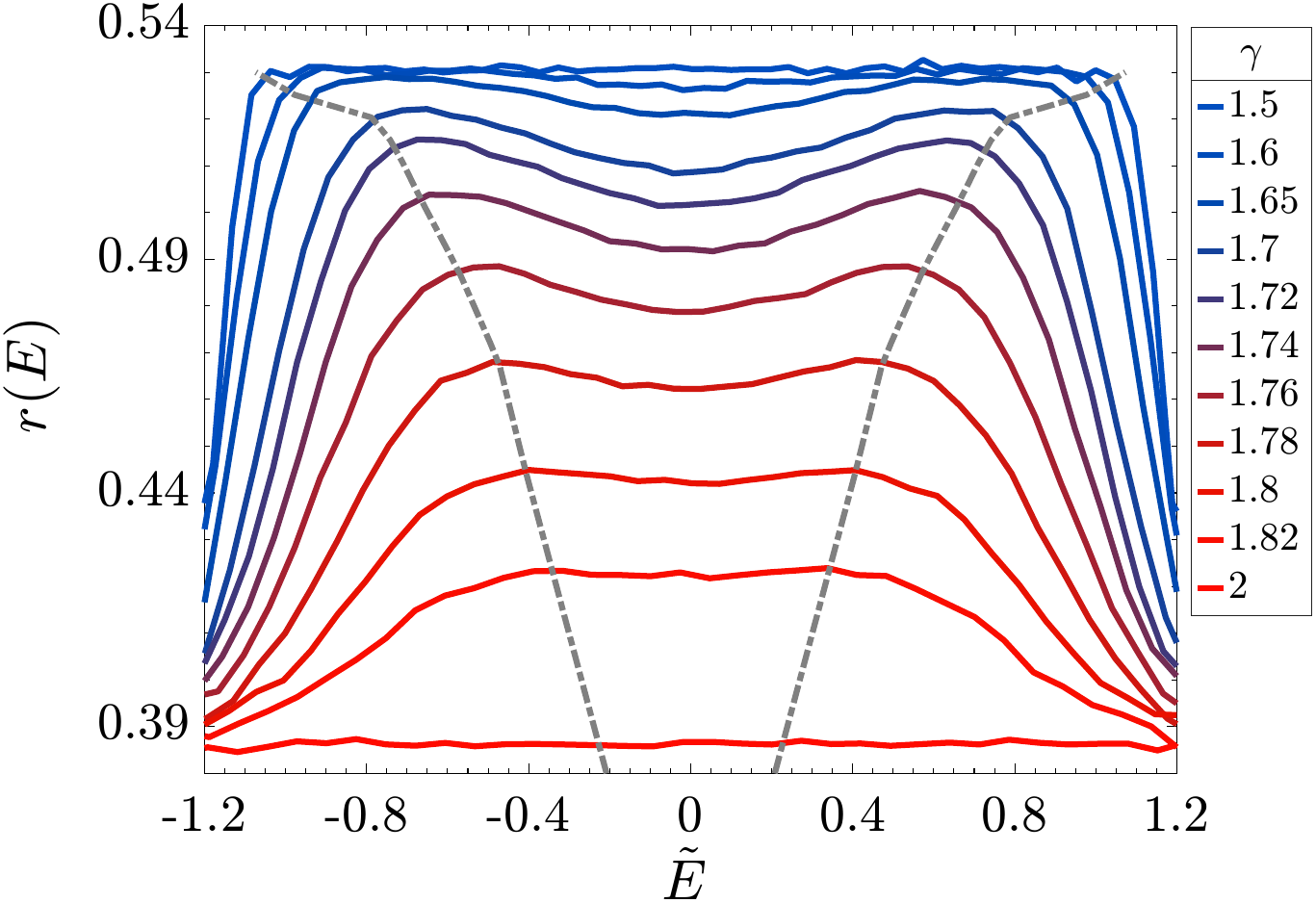}
    \caption[Mobility edge]{{\bfseries Mobility edge.} Ratio of level spacing as a function of energy for $N = 4096$ and different values of $\gamma$. $\tilde{E} = E/2\sgE$ denotes the scaled energy. Gray dashed lines denote the mobility edges.
    }
    \label{fig:r_vs_E}
\end{figure}

The adjacency matrix of the ERG graph~\cite{Alt2021} as well as its weighted version~\cite{Cugliandolo2024} exhibits mobility edge, i.e.~a critical energy separating the localized low-energy eigenstates from the extended excited states. We are interested to know if the addition of on-site disorder affects the presence of the mobility edge in the delocalized phase of sparse random matrices. For localized eigenstates, the energy levels are uncorrelated whereas the extended states hybridize among themselves, producing correlated spectrum. Thus, by looking at the short-range correlation measures as a function of energy, we can identify the mobility edge.

In particular, we look at the energy resolved ratio of level spacing~\cite{Das2023}
\begin{align}
    \begin{split}
    	r(E) &= \frac{1}{N\rho(E)} \sum_{n = 1}^N r_n \delta\del{E - E_n}\\
    	r_n &= \min\del{\tilde{r}_n, \frac{1}{\tilde{r}_n}},\: \tilde{r}_n = \frac{E_{n+1} - E_n}{E_n - E_{n-1}}
    \end{split}
\end{align}
In Fig.~\ref{fig:r_vs_E}, we show the energy dependence of the ratio of level spacing for $N = 4096$ and various values of $\gamma$. The energy axis is scaled as $\tilde{E} = E/2\sgE$ to ensure that the spectral width is independent of $\gamma$. We identify the mobility edge as the energy where $r(E)$ deviates from its bulk behavior. As shown in Fig.~\ref{fig:r_vs_E}, the mobility edges (gray dashed curves) shrink towards zero energy as we increase $\gamma$. For $\gamma > 2$, the ratio of level spacing is homogeneous over the spectrum and reaches the Poisson limit. Thus, we find that despite the addition of the on-site disorder, the mobility edge survives in the delocalized phase of the sparse random matrices.

\subsection{Long-range energy correlations}
Now, we look at the correlation between two energy levels which are far apart from each other. In a series of seminal works, Dyson showed that the energy levels of the Wigner-Dyson ensembles can be interpreted as the equilibrium configuration of Coulomb gas particles confined on a line at a temperature inversely proportional to the degree of level repulsion~\cite{Dyson1962, Dyson1962a}. Thus, for a fixed $N$, upon decreasing the temperature (i.e.~increasing the degree of level repulsion), the Coulomb gas particles freeze to their respective mean positions and produce a picket-fence spectrum~\cite{Pandey1991, Sorathia2012, TorresHerrera2019}. 
As we increase the temperature of the Coulomb gas, in the analogous energy spectrum, the energy levels can fluctuate more around their mean positions and eventually describe a Poisson process where all the energy correlations are lost~\cite{Das2022, Das2023, Roy2025}.

The fluctuation of the $n$th unfolded energy level, $\Ecal_n$ around its mean position, $\mean{\Ecal_n} = n$ can be expressed as
\begin{align}
    \delta_n\equiv \mathcal{E}_n - n = \sum_{k = 1}^{N} c_k \Psi_k(n)
    \label{eq:delta_n_mode}
\end{align}
where $\cbr{\omega_k, \ket{\Psi_k}}$ is an eigenpair of the correlation matrix
\begin{align}
    \mathcal{C}_{mn} = \mean{\Ecal_m \Ecal_n} - \mean{\Ecal_m}\mean{\Ecal_n}.
\end{align}
$\ket{\Psi_k}$ is the $k$th normal mode of the energy level fluctuations with a wavelength $\sim \mathcal{O}(k^{-1})$ and mean square amplitude $\omega_k$~\cite{Andersen1999, Jackson2001}. Then, $\omega_k$ as a function of the wavenumber $k$ gives the dispersion relation of the normal modes. In particular, the normal mode $\ket{\Psi_1}$ with the longest wavelength describes the collective motion where all the energy levels have a common shift without any change in the relative distances. For GOE,  $\omega_k \propto k^{-1}$ and for Poisson ensemble, $\omega_k \propto k^{-2}$ and corresponding normal modes are called rigid and soft respectively. Rigid short (long)-wavelength modes indicate the presence of short (long)-range energy correlations. Thus, the dispersion relation of the energy level fluctuations reflects the nature of energy correlations across all possible length scales.

\subsubsection{Level Number Variance}
To understand the long-range energy correlations in case of the sGOE, we compute the number variance~\cite{Baecker1995, Altshuler1988}
\begin{align}
	\begin{split}
		\Sigma^2(\Egap) &= \mean{ [\mathcal{N}(\Ecal_0, \Egap) - \Egap]^2}_{\Ecal_0}\\
		\mathcal{N}(\Ecal_0, \Egap) &= I\del{\Ecal_0 + \frac{\Egap}{2}} - I\del{\Ecal_0 - \frac{\Egap}{2}}
	\end{split}
	\label{eq:nvar_def}
\end{align}
where $I(\Ecal)$ is the cumulative distribution of the unfolded energy levels, $\mathcal{N}(\Ecal_0, \Egap)$ is the number of energy levels in the window of span $\Egap$ and centered at $\Ecal_0$. 
In case of GOE, the number variance behaves as~\cite{Mehta2004book}
\begin{align}
    \begin{split}
        \Sigma_{\mathrm{GOE}}^2(\Egap) &\approx
    \frac{2}{\pi^2}\left( \ln(2\pi\Egap) + \gEM + 1 -\frac{\pi^2}{8}\right) 
    \end{split}
    \label{eq:nvar_GOE}
\end{align}
where $\gEM = 0.577216\dots$ is the Euler-Mascheroni constant. 
The logarithmic behavior of the number variance (Eq.~\eqref{eq:nvar_GOE}) is a consequence of the \textit{rigidity} of the energy spectrum where the fluctuations of an energy level from its average position over different disorder realizations are restricted due to the correlations with the rest of the spectrum.

In contrast, the number variance of the Poisson ensemble has a linear behavior, $\Sigma^2(\Egap) = \Egap$, indicating the absence of correlation at any length scale.

\begin{figure}
    \includegraphics[width=0.8\linewidth]{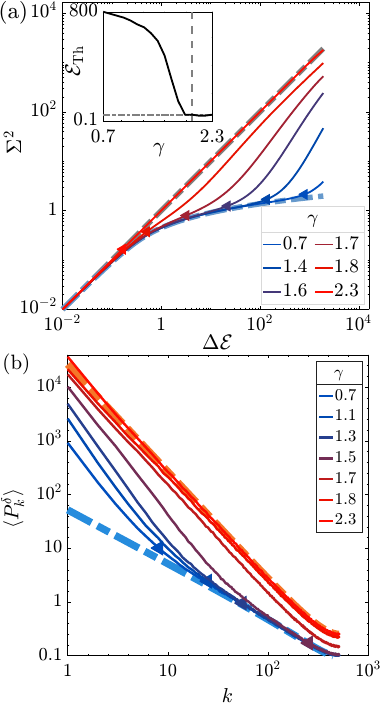}
    \caption[Long-range energy correlation of sGOE]{{\bfseries Long-range energy correlation of sGOE.} (a)~Number variance as a function of energy gap for $N = 16384$ and different values of $\gamma$. Makers denote the unfolded Thouless energy, $\Ecal_\mathrm{Th}$. Dashed blue (gray) curve denotes the analytical expression of GOE (Poisson ensemble) given in Eq.~\eqref{eq:nvar_GOE}. Inset shows the unfolded Thouless energy as a function of $\gamma$.
    (b)~Ensemble averaged power-spectrum of noise for $N = 1024$ and different values of $\gamma$. Dashed blue (red) curve denotes the analytical expression of GOE (Poisson ensemble). Markers denote the Thouless frequency $\kTh = N/\Ecal_\mathrm{Th}$.
    }
    \label{fig:long-range_sRM}
\end{figure}

In various disordered single-particle~\cite{Efetov1983, Cuevas1997, Sierant2020}, many-body~\cite{Bertrand2016, Serbyn2017, GarciaGarcia2018, Schiulaz2019, Corps2020}, random matrix models~\cite{Kravtsov2015, Das2022a, Das2025a, Erdos2015}, there exists a mesoscopic energy scale called the \textit{Thouless energy}, $\Eth$~\cite{Thouless1974, Thouless1975, Thouless1977}, below which the energy levels are correlated similar to the GOE and exhibit universal statistical properties. On the other hand, two energy levels with gap larger than $\Eth$ can be either weakly correlated~\cite{Sivan1987, Aronov1995, Jagannathan2023} or completely uncorrelated such that
\begin{align}
	\Sigma^2(\Egap) \propto \begin{cases}
		\dfrac{2}{\pi^2}\ln \Egap, & \Egap < \Ecal_\mathrm{Th}\\
		\Egap, & \Egap > \Ecal_\mathrm{Th}
	\end{cases}.
	\label{eq:nvar_typ}
\end{align}
Thus, comparing the number variance of an energy spectrum with that of the GOE (Eq.~\eqref{eq:nvar_GOE}), one can estimate the Thouless energy scale. Ref.~\cite{Jackson2001} showed that for $pN \sim \mathcal{O}(1)$, the long-range correlation is homogeneous in the bulk spectrum of sGOE. In Fig.~\ref{fig:long-range_sRM}(a), we show the ensemble averaged number variance as a function of the energy gap for different values of $\gamma$ and observe a crossover from a logarithmic to a linear behavior as we increase $\gamma$.

We identify the unfolded Thouless energy scale via markers beyond which the number variance deviates from $\Sigma_{\mathrm{GOE}}^2(\Egap)$ (Eq.~\eqref{eq:nvar_GOE}) by a threshold value. In the inset of Fig.~\ref{fig:long-range_sRM}(a), we show that the Thouless energy is constant for $\gamma > 2$ and increases monotonically for $\gamma < 2$ as we decrease $\gamma$. 
We observe that for $\gamma\geq 2$ the Thouless energy $\Ecal_\mathrm{Th} \sim \mathcal{O}(1)$ is independent of the system size, implying uncorrelated spectrum across all possible length scales.
 In contrast, the Thouless energy is an extensive quantity for $\gamma < 2$, below which the energy correlations follow the universal features predicted by the random matrix theory.

\subsubsection{Power Spectrum}
We can probe the nature of long-range energy correlation in the frequency space by computing the power-spectrum of the $\delta_n$ statistics~\cite{Faleiro2004, Riser2017}
\begin{align}
	P_k^\delta = |\hat{\delta}_k|^2,\quad \hat{\delta}_k = \frac{1}{\sqrt{N}}\sum_{n}\delta_n\exp\del{-\frac{i2\pi kn}{N}} 
\end{align}
The ensemble averaged power-spectrum exhibits a homogeneous linear  decay w.r.t.~the positive frequencies in case of the GOE whereas the decay is quadratic in Poisson ensembles. However, the power-spectrum exhibits a heterogeneous behavior separated by a critical frequency, $\kTh$, in systems with a well-defined Thouless energy $\Eth$, for e.g.~disordered Heisenberg spin-chain~\cite{Corps2020, Corps2021}, Rosenzweig-Porter ensemble~\cite{Berkovits2020, Das2022}. As the frequency is Fourier dual to the energy, the presence of GOE like energy correlation below $\Eth$ should be reflected in the linear dependence of the high-frequency $\mean{P^\delta_k}$ for $k > \kTh$. On the other hand, the absence of energy correlation above $\Eth$ should lead to a quadratic behavior of the low-frequency power-spectrum for $k < \kTh$. Thus, we expect the following heterogeneous behavior for sGOE
\begin{align}
	\mean{P_k^\delta}\propto \begin{cases}
		k^{-2}, & k < \kTh\\
		k^{-1}, & k > \kTh
	\end{cases}, \quad \kTh \sim \frac{N}{\Ecal_\mathrm{Th}}.
\end{align}
Note that, a complimentary heterogeneous behavior is found in the nonergodic phase of the $\beta$-ensemble~\cite{Das2022, Das2023, Roy2025}, banded random matrix~\cite{Das2025a}, where anomalous long-range energy correlations exist despite the absence of any short-range correlation.

In Fig.~\ref{fig:long-range_sRM}(b), we show the ensemble averaged power-spectrum of noise for $N = 1024$ and different values of $\gamma$ along with the analytical expressions of the GOE and Poisson ensemble~\cite{Faleiro2004}. We find that for $\gamma > 2$, the power-spectrum exhibits a quadratic decay for all frequencies and is well described by the Poisson expression, indicating the absence of energy correlations at all length scales.

On the other hand, for small values of $\gamma$, the power-spectrum follows the GOE expression at high frequencies but deviates from the linear decay below the Thouless frequency (shown by markers in Fig.~\ref{fig:long-range_sRM}(c)). This confirms our conclusions from the number variance, i.e.~for $0 < \gamma < 2$, there is energy correlation below the Thouless energy scale while the long-range correlations are absent.

\section{Discussions}\label{sec_Discussion}
In this work, we investigate a random matrix model called sGOE to show the effects of diagonal disorder in a sparse random matrix ensemble analogous to the adjacency matrix of a weighted Erd\H{o}s-R\'enyi graph. Consistent with earlier studies~\cite{Cugliandolo2024}, we obtain the Anderson localization transition at the critical percolation limit $\gamma=2$ even in the presence of diagonal disorder where the sparsity is parameterized as $p=N^{-\gamma/2}$. The density of states also exhibits a transition from the Wigner semicircle to a Gaussian distribution at $\gamma = 2$. By analyzing the IPRs and the entanglement entropy of the ground state, we observe a multifractal NEE ground state for $0<\gamma<\gamma_0$ while the ground state is localized for $\gamma>\gamma_0$, implying ground state delocalization-localization transition at $\gamma_0 \approx 1.86$.

We analytically calculate the second and fourth energy moments of sGOE and find that the shifted kurtosis exhibits a second order transition at $\gamma=2$. Furthermore, we numerically showe that a mobility edge separating extended and localized eigenstates in the energy spectrum persists even in the presence of diagonal disorder.

An analysis of long-range spectral correlations showed characteristics of typical interacting many-body systems, indicating the presence of a Thouless energy scale $\Eth$. This behavior is also reflected in the power spectrum, which shows distinct scaling regimes above and below a characteristic frequency $k_{\mathrm{Th}}$. For future work, we aim to probe the non-equilibrium dynamics using the spectral form factor and the survival probability in order to identify the relevant timescales in the sparse random matrix ensembles.


\begin{acknowledgements}
    We thank Ivan M.~Khaymovich and Achilleas Lazarides for many useful comments on the manuscript. We acknowledge the support from the Kepler Computing facility, maintained by the Department of Physical Sciences, IISER Kolkata, for various computational needs. A.~K.~D. acknowledges support from the Leverhulme Trust Research Project Grant RPG-2025-063.
\end{acknowledgements}

\appendix
\section{Energy moments of sGOE}\label{sec_apnd_E_mom}
The density in the matrix space (Eq.~\eqref{eq:P_H_sRM}) allows us to obtain the ensemble averaged energy moments. Specifically, $\mean{E^n} = \frac{1}{N}\Tr H^n$ due to cyclic property of the trace~\cite{Bauer2001}. Then,
\begin{align}
    \mean{\mean{E^n}} = \frac{1}{N} \int dH\  \prob{H} \Tr H^n
\end{align}
where $\mean{\mean{\dots}}$ implies an average over the energy spectrum as well as the ensemble of $H$. From the symmetry of $H$ we obtain,
\begin{align}
\begin{split}\label{eq:trH2}
     \mathrm{Tr}\ H^2 = \sum\limits_{i}H_{ii}^2 + 2\sum\limits_{i<j}H_{ij}^2 
\end{split} 
\end{align}
Equation~\eqref{eq:P_Hij_sRM} implies that the diagonal elements are random variables with unit variance and the non-zero off-diagonal elements are random variables with variance $\frac{1}{2}$. Thus, ensemble average of the above trace is
\begin{align}
    \begin{split}
        \expval{\mathrm{Tr} \ H^2} &= \sum\limits_{i}1+2\sum\limits_{i<j}\qty( (1-p) \times 0 + p\times \frac{1}{2})\\
    &= N + 2\times \frac{N(N-1)}{2}\times \frac{p}{2}\\
    &= N + \frac{N(N-1)p}{2}
    \end{split}
\end{align}
Hence, the ensemble averaged 2nd moment of the energy is
\begin{align}
    \expval{E^2} = \frac{1}{N}\qty(N + \frac{N(N-1)p}{2}) = 1+\frac{p}{2}(N-1)
    \label{eq:E2_sRM}
\end{align}
The 4th moment of the energy is $\expval{E^4} = \frac{1}{N}\Tr H^4$. Note that,
\begin{align}\label{trH4}
    \begin{split}
        \Tr H^4 &= \Tr(H^2\cdot H^2) = \sum\limits_{a} (H^2\cdot H^2)_{aa}\\
        &= \sum\limits_{a,b} (H^2)_{ab}(H^2)_{ba}=\sum\limits_{a,b}[(H^2)_{ab}]^2
    \end{split}
\end{align}
For the diagonal elements ($a=b=i$), we get
\begin{align}
    \begin{split}
        \expval{[(H^2)_{ii}]^2} &= \expval{\qty[\sum\limits_k H_{ik}^2]^2} = \sum \limits_k \expval{H_{ik}^4 }+ \sum\limits_{k\neq l} \expval{H^2_{ik}H^2_{il}}\\
        &= \sum \limits_k \expval{H_{ik}^4 }+ \sum\limits_{k\neq l} \expval{H^2_{ik}}\expval{H^2_{il}}
    \end{split}
\end{align}
For the first term, if $k=i$ we have $ \expval{H_{ii}^4 }=3$ from the 4th moment of the Gaussian distribution. Contrarily, if $k\neq i$, we have $N-1$ terms, each having a value $(1-p)\times 0 + p\times \qty(3\cdot \frac{1}{4}) =\frac{3}{4}p$. Note that, the second term can be written as
\begin{align*}
    \sum\limits_{k\neq l} \expval{H^2_{ik}}\expval{H^2_{il}} &= \qty[\sum\limits_k \expval{H_{ik}^2}]^2 - \sum\limits_k \expval{H^2_{ik}}^2\\
    &=\qty[1+ \frac{p}{2}(N-1)]^2 - \qty[1 + (N-1)\frac{p^2}{4}]
\end{align*}
Thus, for the diagonal terms, we get a total contribution of
\begin{align}
    N\qty [3 + (N-1)\frac{3p}{4} + \qty(1 + \frac{p}{2}(N-1))^2 - \qty(1 + (N-1)\frac{p^2}{4})]    \label{eq:E4_term1}
\end{align}
For the off-diagonal terms, $a = i\neq b = l$ we have:
\begin{align}\label{trH4_1}
    \begin{split}
        \expval{[(H^2)_{il}]^2} &= \expval{H^2_{il}H^2_{il}} = \expval{\qty(\sum\limits_{k}H_{ik}H_{kl})\qty(\sum\limits_r H_{ir}H_{rl})}\\
        &= \sum\limits_{k,r} \expval{H_{ik}H_{kl}H_{ir}H_{rl}}
    \end{split}
\end{align}
If $k\neq r$, as $i\neq l$ all terms become independent and vanish, since mean of the entries is zero. So, we consider only $k=r$ case:
\begin{itemize}
    \item $k=r=i$ or $k=r=l$
    \begin{align*}
        \expval{H^2_{ii}H^2_{il}} = 1\times \frac{p}{2}\quad \ \ \ \expval{H^2_{il}H^2_{ll}} = 1\times \frac{p}{2}
    \end{align*}
    \item $k=r\neq i,l$
    \begin{align*}
        \sum_{k\neq i,l}\expval{H_{ik}^2H_{kl}^2} = (N-2)\times \frac{p}{2}\times \frac{p}{2} = (N-2)\frac{p^2}{4}
    \end{align*}
\end{itemize}
Thus, for the off-diagonal terms, we obtain a total contribution of, 
\begin{align}
    \qty[p +(N-2)\frac{p^2}{4}]\times N(N-1)
    \label{eq:E4_term2}
\end{align}
Adding both the contributions from Eqs.~\eqref{eq:E4_term1} and \eqref{eq:E4_term2}, we obtain the ensemble averaged fourth moment of energy for sGOE
\begin{align}
\mean{\expval{E^4}}
= \frac{p^2 N^2}{2}
+ \left(\frac{11}{4}p - \frac{3}{2}p^2\right) N
+ \left(p^2 - \frac{11}{4}p + 3\right)
\label{eq:E4_sRM}
\end{align}

\begin{figure}[t!]
    \centering
    \includegraphics[width=0.8\columnwidth]{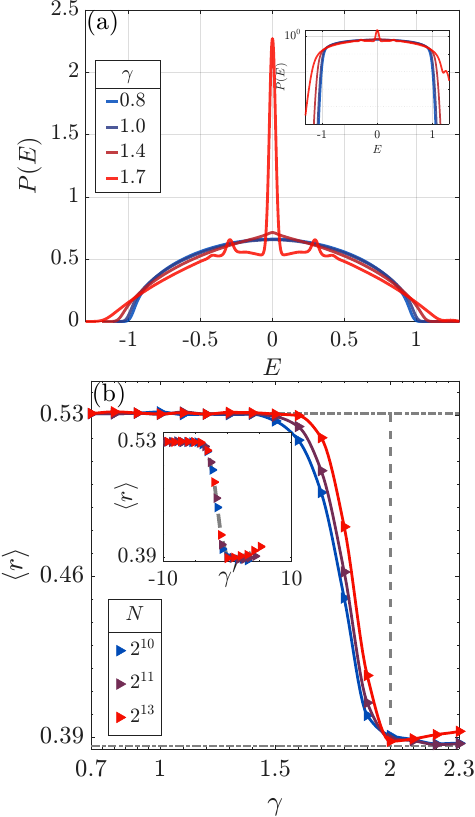}
    \caption{\textbf{Erd\H{o}s-R\'enyi-Gilbert graph}. (a)~Ensemble-averaged density of States for $\mathcal{A}(N, p)$ with $N=1024$ and different values of $\gamma$ ($p= N^{-\gamma/2}$). The energy variance is fixed to be $\sgE = \frac{1}{4}$. Inset shows the DOS in log-scale.
    (b)~Ratio of level spacing vs.~$\gamma$ for different system sizes. Inset shows the collapsed data following 2nd order transition ansatz (Eq.~\eqref{eq:2nd_ansatz}) with $\gamma_c=2$ and $\nu \approx 0.94$.
    }
    \label{fig:ERG_DOS}
\end{figure}

\section{Statistical properties of the ERG graph}\label{app:ERG_stat}

{
In this section, we show the statistical properties of $\mathcal{A}(N,p)$, the adjacency matrix of the ERG graph with $N$ number of nodes and $p$ is the probability of two nodes being connected. Following the same parameterization as sGOE, we take $p= N^{-\gamma/2}$.\\[0.2cm]

Since $\mathcal{A}_{ij}\in \{0,1\}$, we have \begin{equation}\label{eq:P_Aij_ERG}
    \mathrm{P}(\mathcal{A}_{ij}) = (1-p) \delta(\mathcal{A}_{ij}) + p\delta(\mathcal{A}_{ij}-1)
\end{equation}
and in a similar way as Eq. \eqref{eq:P_H_sRM} we define the probability distribution of the adjacency matrix, 
\begin{equation}\label{eq:P_A_ERG}
     \mathrm{P}(\mathcal{A}) = \prod\limits_{i<j}  \mathrm{P}(\mathcal{A}_{ij}) = \prod\limits_{i<j} \qty[(1-p) \delta(\mathcal{A}_{ij}) + p\delta(\mathcal{A}_{ij}-1)]
\end{equation}
}
In Fig.~\ref{fig:ERG_DOS}, we show the DOS of $\mathcal{A}(N,p)$ for different values of $\gamma$ and $N = 1024$. For small values $\gamma$, the DOS converges to Wigner's semicircle law, Eq.~\eqref{eq:DOS_SC} as observed in case of GOE. For larger values of $\gamma$, there are extensive number of isolated vertices supporting localized eigenstates at zero energy \cite{Evangelou1992}. Consequently, the DOS develops a divergence at zero energy, see Fig.~\ref{fig:ERG_DOS} for $\gamma = 1.7$. Such degenerate zero modes span Fock space cages disconnected from the rest of the Hilbert space as also observed in systems with chiral symmetry~\cite{Nicolau2026, Jonay2026}. Importantly, addition of diagonal disorder in case of sGOE removes these zero modes from the spectrum and DOS exhibit no divergence at zero energy, see Fig.~\ref{fig:E_mom_sRM}.

To understand the energy correlation of $\mathcal{A}(N,p)$, we discard the zero modes and compute the ratio of level spacing, as shown in Fig.~\ref{fig:ERG_DOS} for different system sizes. We show that the crossover curves of the mean ratio of level spacing as a function of $\gamma$ can be collapsed following the ansatz in Eq.~\eqref{eq:2nd_ansatz} with the critical point being $\gamma \approx 2$. Thus, short-range energy correlation of $\mathcal{A}(N,p)$ exhibit a 2nd order transition at the critical percolation limit $p = N^{-1}$. This is consistent with the delocalization-localization transition at $p = N^{-1}$ in case of the adjacency matrix of the ERG graph~\cite{Alt2021}, i.e.~the neighboring bulk energy levels are correlated (uncorrelated) for $p > N^{-1}$ ($p < N^{-1}$), similar to sGOE.

{
\subsection{Energy Moments of ERG}\label{ERG_gs}
\begin{figure}
    \centering
    \includegraphics[width=0.86\linewidth]{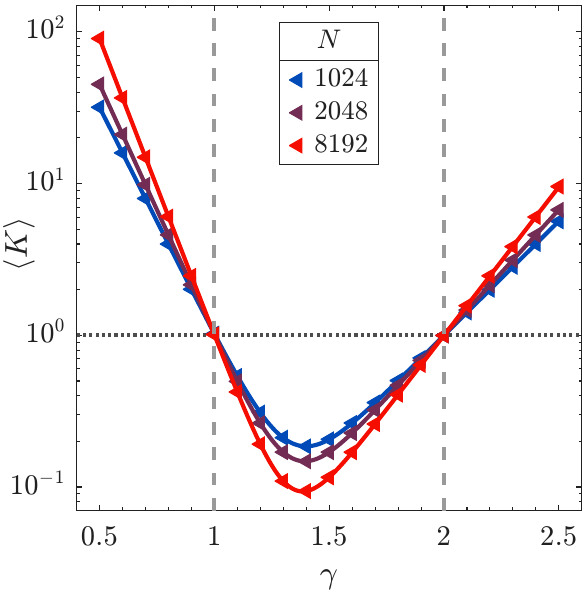}
    \caption{\textbf{Kurtosis for ERG graph.} Shifted kurtosis for the ERG graph for $N=1024,2048,8192$. The markers denote numerical result while the solid lines denote the analytical result using Eq. \eqref{app:en_kurt}. The vertical lines denote $\gamma=1,2$ while the horizontal line denotes $K=1$.}. 
    \label{fig:ERGkurt}
\end{figure}

Similar to Appendix \ref{sec_apnd_E_mom}, we calculate the energy moments for the adjancency matrix of the ERG graph. Note that $\expval{A_{ij}} = 1\times p + 0\times (1-p) = p$ and hence unlike the sGOE, the odd energy moments need not be zero. Also we have $(\mathcal{A}_{ij})^n = \mathcal{A}_{ij} \implies \expval{(\mathcal{A}_{ij})^n} = p$ for all $n \geq 1$. Since $\Tr \mathcal{A} = 0$ (no self-loops) we have $\expval{E}=0$. As $\mathcal{A}_{ii} =0$, from Eq. \eqref{eq:trH2} we have the second energy moment, 
\begin{equation}
    \expval{E^2} = \frac{1}{N} \qty[2 \sum\limits_{i<j}\expval{\mathcal{A}_{ij}^2}] = p(N-1)
\end{equation}
For $\expval{E^3}$ we have to calculate $\expval{\Tr \mathcal{A}^3}$. 
\begin{equation}
    \expval{\Tr \mathcal{A}^3} = \sum\limits_m \expval{(\mathcal{A}^3)_{mm}} = \sum\limits_{m}\sum\limits_{n}\sum\limits_{p} \expval{\mathcal{A}_{mn}\mathcal{A}_{np}\mathcal{A}_{pm}}
\end{equation}
Note that a term in the above sum is non-zero only if all the three indices are pairwise distinct, since the equality of any one index will result in a diagonal term of $\mathcal{A}$ in the expectation, which is zero. If all the indices are distinct, the expectation value yields $p^3$. Since the indices are ordered, the number of distinct triples is simply given by $\nPr[N]{3} = N(N-1)(N-2)$. Thus the third energy moment is,
\begin{equation}
    \expval{E^3} = \frac{1}{N}\qty[N (N-1)(N-2) p^3] = (N-1)(N-2)p^3
\end{equation}
Finally, for $\expval{E^4}$ we have to calculate $\expval{\Tr \mathcal{A}^4}$. We start with Eq. \eqref{trH4} in Appendix \ref{sec_apnd_E_mom}. For the diagonal terms $a=b=i$ we have:
\begin{align}
    \begin{split}
        \expval{[(\mathcal{A}^2)_{ii}]^2} &= \expval{\qty[\sum\limits_k \mathcal{A}_{ik}^2]^2} = \sum \limits_k \expval{\mathcal{A}_{ik}^4 }+ \sum\limits_{k\neq l} \expval{\mathcal{A}^2_{ik}\mathcal{A}^2_{il}}\\
        &= \sum \limits_k \expval{\mathcal{A}_{ik}^4 }+ \sum\limits_{k\neq l} \expval{\mathcal{A}^2_{ik}}\expval{\mathcal{A}^2_{il}}\\
        &= \sum \limits_k \expval{\mathcal{A}_{ik} }+ \sum\limits_{k\neq l} \expval{\mathcal{A}_{ik}}\expval{\mathcal{A}_{il}}
    \end{split}
\end{align}
The first term is non-zero when $k\neq i$  ($(N-1)$ such terms), giving us $p(N-1)$. Similar to the sGOE calculation, the second term can be written as $[p(N-1)]^2 - (N-1)p^2$. Therefore the diagonal contribution becomes, 
\begin{align}
    N\Bigg[p(N-1) + &p^2(N-1)^2 - p^2(N-1) \Bigg] \\ &= N\qty[p(1-p)(N-1) + p^2(N-1)^2]
\end{align}
For the off-diagonal term $a=i \neq b =l$, we refer to Eq. \eqref{trH4_1}. Unlike sGOE, the case where $k\neq r$ will contribute, since $\expval{\mathcal{A}_{ij}}=p \neq 0$. In this case, we have $(N-2)(N-3)$ terms, each having a value $p^4$.\\[0.2cm]
When $k=r$, then the case $k=r=i$ or $k=r=l$ will yield zero due to the presence of the diagonal term. The case where $k=r \neq i,l$ gives us $(N-2)p^2$. Thus the total off-diagonal contribution is, 
\begin{equation}
    N(N-1)\qty[p^2(N-2)+(N-2)(N-3)p^4]
\end{equation}
Adding both the contributions, we obtain the fourth energy moment 
\begin{equation}
   \expval{E^4} = p(N-1)(1+p(N-2)(p^2(N-3)+2))
\end{equation}
As before, we calculate the shifted kurtosis, 
\begin{equation}
    K = \frac{(1-2p) + p^3 (N-2)(N-3)}{p(N-1)}\label{app:en_kurt}
\end{equation}
In the limit $N\gg 1$ we have,
\begin{equation}
    K \to \frac{1}{pN} +p^2N = N^{1-\gamma} + N^{\gamma/2-1}
\end{equation}
At $\gamma=1$ we have $K\to 1+N^{-1/2}$ and at $ \gamma=2$ we have $K\to 1 + N^{-1}$.
}

\subsection{Ground State of ERG}

\begin{figure}
    \includegraphics[width=\linewidth]{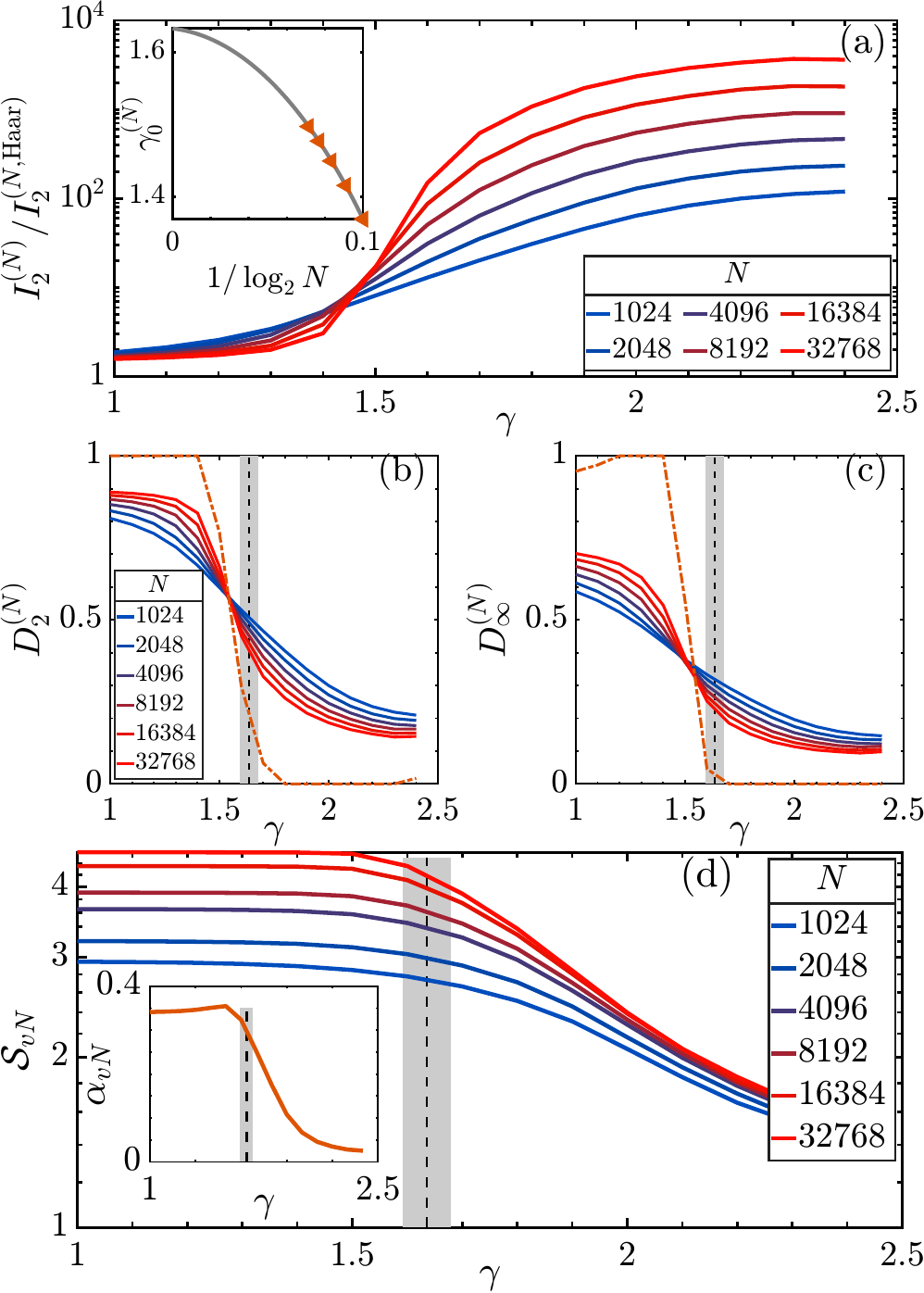}
    \caption{{\bfseries Ground state of Erd\H{o}s-R\'enyi-Gilbert graph.} (a)~Inverse participation ratio scaled by that of the Haar random states for different system sizes. Inset shows the system size dependence of the intersection points of consecutive crossover curves, indicating that that critical point is $1.6353 \pm 0.0868$.
    Finite size fractal dimensions (b)~$D_2^{(N)}$ and (c)~$D_\infty^{(N)}$ as a function of $\gamma$ for different values of $N$. Red dashed lines denote the thermodynamic extrapolations, $D_2$ and $D_\infty$.
    (d)~bipartite von Neumann entanglement entropy for different system sizes. Inset shows the system size scaling, $\mathcal{S}_{vN} \propto \alpha_{vN} \ln N$. In panels (b), (c) and (d), vertical dashed line denotes the delocalization-localization transition point for ground state and corresponding shaded area denotes 95\% confidence interval.
    }
    \label{fig:ERG_GS}
\end{figure}

Having understood the bulk spectral properties, we now look at the ground state of $\mathcal{A}(N,p)$. In Fig.~\ref{fig:ERG_GS}(a), we show the ensemble averaged IPR, $I_2^{(N)}$ (Eq.~\eqref{eq:def_Iq}) of the ground state scaled by that of the Haar random states, $I_2^{(N, \mathrm{Haar})} = 2/(N+1)$. In the inset of Fig.~\ref{fig:ERG_GS}(a), we show $\gamma_0^{(N)}$, the intersection points of the consecutive crossover curves as a function of $1/\log_2 N$ and a quadratic fit indicates that the critical point in the thermodynamic limit ($N\to \infty$) is $\gamma_0 = 1.6353 \pm 0.0868$. Thus, the ratio of the ensemble averaged IPR of the ground state and Haar random state converges to unity for $\gamma \lessapprox \gamma_0$ and diverges with system size $N$ for $\gamma \gtrapprox \gamma_0$. This indicates that the ground state of $\mathcal{A}(N,p)$ is ergodic for $\gamma \lessapprox \gamma_0$. 

\begin{figure}[!ht]
    \centering
    \includegraphics[width=0.88\columnwidth]{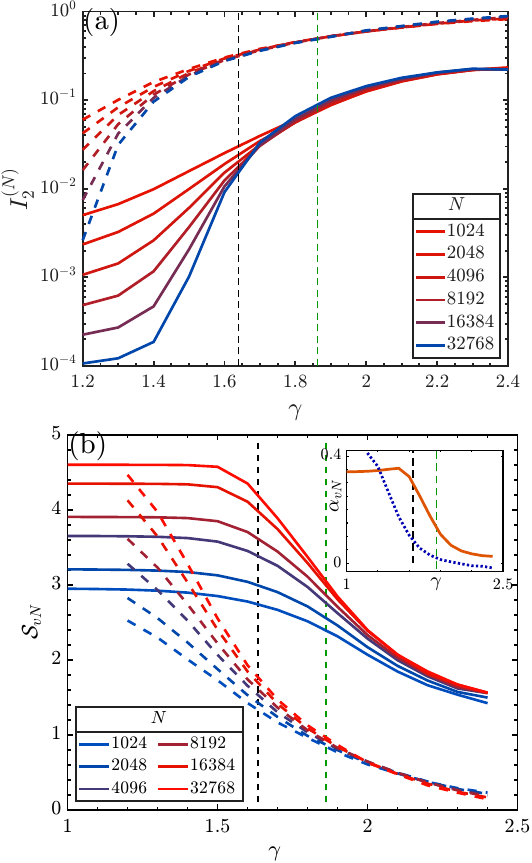}
    \caption{\textbf{Comparison between sGOE and ERG.} (a) Ensemble averaged IPR and (b) Entanglement entropy, vs. $\gamma$, for different values of $N$. In both figures, solid (dashed) lines correspond to ERG (sGOE). The black and green dashed vertical lines denote the zero temperature delocalization-localization transitions at $\gamma_0\approx 1.63$ and $\gamma_0\approx 1.86$ for ERG and sGOE, respectively.}
    \label{fig:sGOE_ERG_compare}
\end{figure}

To verify the above conclusion from the system size scaling of the fluctuations of the ground state intensities, we show the finite size fractal dimensions, $D_2^{(N)}$ and $D_\infty^{(N)}$ vs.~$\gamma$ for different values of $N$ in Figs.~\ref{fig:ERG_GS}(b) and (c), respectively. We find that the crossover curves of finite size fractal dimensions intersect around $\gamma \approx \gamma_0$. In Figs.~\ref{fig:ERG_GS}(b) and (c), we also show the thermodynamic extrapolations of $D_2^{(N)}$ and $D_\infty^{(N)}$ indicating that $D_2 = D_\infty \approx 1$ for $\gamma \lessapprox \gamma_0$. In Fig.~\ref{fig:ERG_GS}(d), we show the bipartite von Neumann entanglement entropy of the ground state as a function of $\gamma$ for different system sizes. The inset of Fig.~\ref{fig:ERG_GS}(d) show that $\mathcal{S}_{vN} \approx \log_2 N / 2$ for $\gamma \lessapprox \gamma_0$, i.e.~the entanglement entropy of the ground state follows the Page value for $\gamma \lessapprox \gamma_0$. Thus, in case of the adjacency matrix of the ERG graph, ground state delocalization-localization transition occurs around $\gamma_0 = 1.6353 \pm 0.0868$ which is distinct from the Anderson transition in the bulk ($\gamma = 2$).

This is in contrast with sGOE where the ground state delocalization-localization transition occur at $\gamma_0 \approx 1.86$. In Fig.~\ref{fig:sGOE_ERG_compare}, we show the ensemble averaged IPR and entanglement entropy of the ground state of sGOE and the adjacency matrix of ERG graph for different system sizes, which show that for a fixed value of $\gamma$ and system size, addition of quenched disorder decreases the correlation length and entanglement of the ground state.

\FloatBarrier
\bibliography{ref_S_RM}

\end{document}